\documentclass[aps,prb,preprint,groupedaddress,floatfix]{revtex4}
\usepackage{graphicx,amssymb}
\begin{document}

\title{Free energy calculations along entropic pathways: I. Homogeneous vapor-liquid nucleation for atomic and molecular systems.} 
\author{Caroline Desgranges and Jerome Delhommelle}
\affiliation{Department of Chemistry, University of North Dakota, Grand Forks ND 58202}
\date{\today}

\begin{abstract}
Using the entropy $S$ as a reaction coordinate, we determine the free energy barrier associated with the formation of a liquid droplet from a supersaturated vapor for atomic and molecular fluids. For this purpose, we develop the $\mu VT-S$ simulation method that combines the advantages of the grand-canonical ensemble, that allows for a direct evaluation of the entropy, and of the umbrella sampling method, that is well suited to the study of an activated process like nucleation. Applying this approach to an atomic system such as $Ar$ allows us to test the method. The results show that the $\mu VT-S$ method gives the correct dependence on supersaturation of the height of the free energy barrier and of the size of the critical droplet, when compared to predictions from classical nucleation theory and to previous simulation results. In addition, it provides insight into the relation between entropy and droplet formation throughout this process. An additional advantage of the $\mu VT-S$ approach is its direct transferability to molecular systems, since it uses the entropy of the system as the reaction coordinate. Applications of the $\mu VT-S$ simulation method to $N_2$ and $CO_2$ are presented and discussed in this work, showing the versatility of the $\mu VT-S$ approach.
\end{abstract}

\maketitle

\section{Introduction}
In recent years, computer simulations have considerably advanced our knowledge on phase transitions and molecular assembly processes by shedding light on the microscopic mechanisms underlying these phenomena~\cite{yasuoka1998molecular,oxtoby1992homogeneous,shen1999computational,weakliem1993toward,schenter1999dynamical,zeng1991gas,yi2002molecular,kinjo1999computer,toxvaerd2001molecular,ford1996thermodynamic,talanquer1995density,reiss1990molecular,kalikmanov1995semiphenomenological,horsch2008modification,neimark2005birth,oxtoby1988nonclassical,lutsko2008density,wang2008homogeneous,ten1999numerical,gonzalez2015bubble,loeffler2015improved,sosso2016crystal,xu2015effect,keasler2015understanding,wilhelmsen2015influence,van2015mechanism,hale1986application,hale2005temperature,hale2010scaled,yuhara2015nucleation,lauricella2015clathrate,singh2014characterization,ni2013effect,reinhardt2014effects}. Simulations have allowed to identify key parameters for these activated processes, which allow to measure and analyze the progress of the system towards the formation of a new phase. These parameters, which are often complex functions of e.g. the coordinates of the atoms or of some features of their collective behavior, are known as order parameters or reaction coordinates (RCs)~\cite{ten1998computer,chen2001aggregation,oh1999formation,chen2002simulating,zhukhovitskii1995molecular,nishi2015molecular,lupi2016pre,santiso2015general,berryman2016early,zimmermann2015nucleation,lam2015atomistic,kratzer2015two,bolhuis2015practical}. RCs are generally tailored to a type of system or to a type of phenomenon under study. This is the case, for instance, for vapor-liquid nucleation, a process of great interest for many applications such as for atmospheric nucleation. Possible choices of RCs are e. g. the radius of the incipient droplet or its size (in terms of numbers of atoms within the droplet)~\cite{ten1998computer,tanaka2005tests,kraska2006molecular,oh2000small,senger1999molecular,lau2015water,toxvaerd2016nucleation}, and the analysis of the underlying pathway is generally made through the determination of the free energy barrier of nucleation and of the critical size of the nucleus beyond which spontaneous growth occurs.

The entropy of the system $S$ is an especially appealing choice for a RC since it is, by definition, directly connected to the onset of order (i.e. $S$ decreases as order increases). It also has the additional advantage of being very general and, as such, can be used for a wide range of systems and phenomena, regardless of the geometry, structure and phases involved. In the first part of this series, our goal is to establish the use of $S$ as a RC for the homogeneous vapor-liquid nucleation of atomic and molecular systems. For this purpose, we develop a simulation method that allows to vary the entropy of the system and, as a result, to drive the formation of the liquid droplet. We then calculate the free energy barrier associated with this process. The simulations carried out in this work therefore allow us to identify the entropic pathway underlying the nucleation process, which starts with the metastable supersaturated vapor and ends at the top of the free energy barrier, when a liquid droplet of a critical size has formed. By varying the conditions of nucleation, we are also able to analyze the impact of the supersaturation on the nucleation process and to rationalize the interplay between the free energy of nucleation, the size of the liquid droplet and the range of entropies spanned during the nucleation process. The validity of the results is assessed through a comparison with the predictions from the classical nucleation theory~\cite{mcgraw1996scaling,mcgraw1997interfacial}, which provides a relation between the free energy of nucleation and the supersaturation, and with prior simulation work~\cite{ten1998computer,koga1999thermodynamic}. Results are presented for both atomic ($Ar$) as well as molecular ($N_2$ and $CO_2$) systems to demonstrate the versatility and the transferability of the method to molecular systems.

The paper is organized as follows. In the next section, we present the simulation method and discuss how the entropy of the system is used as a RC. We also explain how we calculate the free energy profile for the the nucleation process. We then present the molecular force fields used to model the three systems studied in this work, before discussing the technical details. In particular, we give an account of how we obtain and choose the input parameters for the simulations to control the extent of supersaturation during the nucleation process. Then, we discuss the results obtained here on the three systems considered in this work, $Ar$, $N_2$ and $CO_2$ and focus on quantifying the impact of supersaturation on the height of the free energy barrier, as well as on the size of the liquid droplet and on the entropy of the system at the top of the free energy barrier. We finally draw the min conclusions from this work in the last section.

\section{Simulation Method}

\subsection{Sampling entropic pathways}

The $\mu VT-S$ approach is based on the grand-canonical ensemble, where $\mu$ is the chemical potential, $V$ the volume, $T$ the temperature and $S$ the entropy of the system. Having $\mu$ as an input parameter for the simulations leads to a direct comparison with the theoretical predictions from the classical nucleation theory (CNT)~\cite{mcgraw1996scaling,mcgraw1997interfacial}. CNT analyzes the nucleation process, and the free energy barrier that the system has to overcome during this process, as a competition between two contributions, which cancel out when a droplet of a critical size has formed, i.e. when the system has reached the top of the free energy barrier. The first contribution, the surface term, is energetically unfavorable and is the predominant term for small droplets. It results from the cost of creating the interface between the liquid droplet and the vapor and can be evaluated as the product of the area of the incipient spherical droplet by the surface tension. The second contribution, the volume term, is energetically favorable and becomes predominant for larger droplets. The volume term corresponds to the energetic gain in 'converting' the parent phase, the metastable supersaturated vapor, into a droplet of the stable phase, the liquid phase. In the CNT approach~\cite{mcgraw1996scaling,mcgraw1997interfacial}, the volume term, and hence the free energy of nucleation, is proportional to the supersaturation $\Delta \mu=\mu -\mu_v$, which is the difference between the chemical potential of the liquid ($\mu$) and the chemical potential of the supersaturated vapor $\mu_v$ at the same pressure as the liquid. Carrying out simulations in the grand-canonical ensemble allows us to choose $\mu$ and hence the supersaturation $\Delta \mu$ for the nucleation event.

Working in the grand-canonical ensemble also has additional advantages. Since the number of atoms/molecules is allowed to fluctuate, this ensemble provides a direct acces to thermodynamic quantities that are difficult to calculate in other ensembles, as e.g. in the $NVT$ or $NPT$ ensemble.  This is the case, for instance, for the entropy $S$ of the system. In the grand-canonical ensemble, in which the chemical potential $\mu$, the temperature $T$ and volume $V$ are fixed during the simulations, it is straightforward to evaluate $S$ during the simulations through 
\begin{equation}
S={\bar U - \mu \over T}
\label{Sdef}
\end{equation}
where $\bar U$ is the internal energy per atom/molecule, obtained by calculating the potential energy due to the interactions between atoms/molecules and by adding the kinetic (ideal gas) contribution to the internal energy of $k_BT/2$ per degree of freedom.

This means that, in the grand-canonical ensemble, $S$ can be used as the reaction coordinate to measure the onset of order in the system and, in the case of nucleation, its progress towards the formation of a liquid droplet of a critical size. To achieve this, we carry out $\mu VT-S$  simulations by taking advantage of the well-known umbrella sampling method~\cite{torrie1977nonphysical} to allow the system to follow the RC, in this case $S$. From a practical standpoint, we add a bias potential energy, function of the entropy of the system, $U_{bias}(S)$, to the total potential energy of the system. In this work, we use the following harmonic function of $S$ as the bias potential energy
\begin{equation}
U_{bias}= {1\over 2} k (S-S_0)^2 
\label{biasfunc}   
\end{equation}
where $k$ is a spring constant and $S_0$ is the target value for the entropy of the system.

$\mu VT-S$ simulations are carried out within the Monte Carlo (MC) framework, with the usual Metropolis criteria used to accept/reject the different types of MC moves attempted on the atoms/molecules of the system. More specifically, since these simulations are rooted in the grand-canonical ensemble, the conventional Metropolis criteria, as obtained e.g. by Allen and Tildesley~\cite{Allen} in the grand-canonical ensemble, are used in $\mu VT-S$ simulations. For instance, for the MC steps involving the translation of single atom/molecule, as well as for the rotation of a molecule, we use the following Metropolis criterion
\begin{equation}
acc(o \to n)=min\left[ 1, {\exp (-\beta \Delta U} \right]
\label{Metro_trans}   
\end{equation}
with $\Delta U$ denoting the change in the total potential energy (interaction energy+bias potential) corresponding to the move from the old ($o$) configuration to the new ($n$) configuration. Similarly, for the MC moves involving the insertion of a new atom/molecule, we use the following criterion:
\begin{equation}
acc(o \to n)=min\left[ 1, {\exp (-\beta \Delta U)+ \ln \left({ N \over {zV}} \right) }  \right]
\label{Metro_ins}   
\end{equation}
where $z$ is the activity $z=\exp(\beta \mu) / \Lambda^3$, where $\Lambda$ is the de Broglie wavelength and $N$ is the current number of atoms/molecules in the system. In the case of molecular fluids, one also needs to take into account the terms related to the other degrees of freedom. For linear molecules like $N_2$ and $CO_2$, we include for the rotation a factor of $8 \pi I k_BT / 2h^2$, where $I$ is the moment of inertia of the linear molecule, $k_B$ is the Boltzmann constant and $h$ is Planck's constant. 

For all systems ($Ar$, $CO_2$ and $N_2$), we carry out $\mu VT-S$ simulations in cubic cells with an edge of $100$~\AA~and apply the usual $3D$ periodic boundary conditions. To simulate the entire nucleation process, we carry out a series of umbrella sampling simulations with decreasing values for the target entropy $S_0$.  This allows us to sample the underlying entropic pathway, which goes from the high entropy parent phase (supersaturated vapor) to the low entropy phase (liquid). During the course of the simulation, we collect histograms for the number of times each entropy interval is visited. These histograms are then used to build the free energy profile of nucleation using the techniques developed to analyze the results from umbrella sampling simulations (more details can be found in previous work~\cite{torrie1977nonphysical,Allen,Alu2,Au,desgranges2014unraveling}). For each value of the target entropy $S_0$, simulations are first run for $100 \times 10^6$ $MC$ steps to allow the system to relax. Then, a second run, the production run, is performed for $200 \times 10^6$ $MC$ steps during which the data are collected and the calculation of average properties is carried out. The different types of $MC$ steps are attempted with the following rates. In the case of of $Ar$, $75$~\% of the $MC$ steps consist in the translation of a single atom, while the $25$~\% remaining moves are equally split between the $MC$ steps corresponding to the insertion and deletion of Ar atoms. In terms of computational efficiency, working in the grand-canonical ensemble allows the number of atoms to vary, which means that the simulations for the higher $S$ values will be performed on systems with few atoms. This is advantageous when compared to methods based on the NPT ensemble, which simulate systems with a constant number of atoms. Furthermore, calculating the reaction coordinate $S$ through Eq.~\ref{Sdef} after every time step is very fast. The CPU time for a $\mu VT-S$ run is therefore very close to a conventional grand-canonical run. For instance, a production run carried out on systems containing an Ar droplet of a critical size (i.e. the largest systems sizes during nucleation) takes 30 CPU hours (CPU time given for a single Intel Xeon processor clocked at 3.2 GHz). For $CO_2$ and $N_2$, we take into account the $MC$ steps for the rotation of a single molecule ($37.5$~\%), in addition to the translation ($37.5$~\%) and the insertion/deletion ($25$~\%). Throughout the simulations, we check that we have sufficiently large acceptance rates for the insertion/deletion steps to ensure an efficient sampling. For all systems and all target entropies $S_0$, we obtain acceptance rates greater than $40$~\% for the insertion/deletion steps. 

\subsection{Models}
We model Argon with a Lennard-Jones (LJ) potential
\begin{equation}
\phi(r)=4\epsilon \left[ {\left({\sigma \over r }\right) ^{12}- \left({\sigma \over r }\right) ^6} \right]
\end{equation}
where $\epsilon=117.05~K$ is the depth of the potential well, $\sigma=3.4$~\AA~is the atom diameter and $r$ is the distance between two interacting atoms. The interactions between $Ar$ atoms are calculated for all distances up to $13.6$~\AA~and neglected beyond that cutoff distance. We use this large cutoff distance ($4 \sigma$) and do not apply tail corrections, as in previous work on vapor-liquid nucleation~\cite{ten1998computer}. We add that an alternative approach, relying on the calculation of long-range corrections as a function of the local density, has been developed in recent years for inhomogeneous systems~\cite{ghoufi2016computer}.

For $N_2$ and $CO_2$, we consider both molecules to be rigid and model the interactions with a distribution of point charges to account for the molecular quadrupole and with a distribution of LJ sites to model the dispersion-repulson interactions. For $N_2$, we consider two LJ sites per $N_2$ molecule (one on each atom) and three point charges (one negative charge on each atom and one positive charge at the center of the $N-N$ bond) to model the quadrupole of the $N_2$ molecule. We use the following parameters~\cite{delhommelle2000etablissement} for the LJ sites, $\epsilon_{NN} / k_B = 36$~K and $\sigma_{NN} = 3.30$~\AA, and for the point charges, $q_{center}=0.966~e$ and $q_N=-q_{center}/2$. The $N-N$ bondlength is set to $0.549$~\AA, and use a spherical cutoff set to $4 \sigma_{NN}=13.2$~\AA~for the LJ part of the potential and to $15$~\AA~for the quadrupole-quadrupole interactions. 

For $CO_2$, we use the TraPPE force field~\cite{Potoff} which is based on a distribution of three LJ sites (located on the atoms of the molecule) and of three atomic charges, with the following parameters: $\epsilon_{CC} / k_B = 27~K$,$\epsilon_{OO} / k_B = 79~K$, $\sigma_{CC} = 2.80$~\AA~and $\sigma_{OO} = 3.05$~\AA, $q_c=0.7~e$ and $q_O=-0.35~e$. The parameters for the LJ interactions between unlike atoms are given by the Lorentz-Berthelot mixing rules. The bondlength is set to $1.16$~\AA~as in the original paper~\cite{Potoff}. We use a spherical cutoff set to $4 \sigma_{OO}=12.2$~\AA~for the LJ part of the potential and to $14.6$~\AA~for the quadrupole-quadrupole interactions. 

\subsection{Setting up the simulations}

\subsubsection{Argon}

$\mu VT-S$ simulations for $Ar$ are carried out at $T=128.76$~K. We present in Table~\ref{muAr} the conditions for the three supersaturations studied in this work. They are obtained through the Expanded Wang-Landau (EWL) simulation method we recently developed~\cite{PartI,PartII,PartIII,PartIV} (other methods can also be used to determine this information~\cite{Camp,Pana,Potoff,nezbeda1991new,Singh,rai2007pressure,Rane,expanded,Shi1,eslami2007molecular,Vogt,widom1963some,CBMC}). EWL simulations provide the grand-canonical partition function and the probability distribution for the number of $Ar$ atoms in the system for any value of $\mu$. From there, the value of the chemical potential at the vapor-liquid coexistence can be directly obtained by finding the value of $\mu$ leading to equal probabilities for the vapor and liquid phases (more details are provided in previous work~\cite{PartI}). The EWL simulations also give access to all other thermodynamic properties, including the pressure $P$ and entropies at coexistence, $S_{l}$ and $S_{v}$, also given in Table~\ref{muAr}. This provides an estimate for the range of entropies that needs to be sampled during the $\mu VT-S$ simulations to cover the entire vapor$\to$liquid transition. We also give in Table~\ref{muAr} the supersaturation $\Delta \mu$ that we apply to obtain the supersaturated vapors, which serve as parent phases for the nucleation processes. Increasing the value of the chemical potential by $\Delta \mu$ ($\Delta \mu > 0$) pushes the system further into the domain of the phase diagram where the liquid phase is stable. We add that $\mu_v$ that appears in the definition of the supersaturation $\Delta \mu$ is very close to $\mu_{coex}$ for the low pressures studied here (of the order of $10^{-3}$~kJ/kg). Here, we consider 3 different supersaturations and label each set of conditions as system $1$, $2$ and $3$. 

\begin{table}[hbpt]
\caption{$Ar$ at $128.76$~K: Input parameters for the three supersaturations studied in this work (systems 1, 2 and 3).}
\begin{tabular}{{|c|c|c|c|c|c|c|}}
\hline
$ $ & $\mu$ & $\Delta \mu$  & $P$ & $P/P_{coex}$ & $S_l$ & $S_v$ \\
$ $ & $(kJ/kg)$ & $(kJ/kg)$ & $(bar)$ & $ $ & $(kJ/kg/K)$ & $(kJ/kg/K)$  \\
\hline
\hline
 coex & -300.19 & 0.00 & 21.58 & 1.0 & 1.856 & 2.701  \\
\hline
 system~1 & -298.12 & 2.07 & 43.16 & 2.0 & 1.834 & -  \\
\hline
 system~2 & -297.72 & 2.47 & 47.48 & 2.2 & 1.831 & - \\
 \hline
 system~3 & -297.35 & 2.84 & 51.80 & 2.4  & 1.830 & - \\
\hline
\hline
\end{tabular}
\label{muAr}
\end{table}

\begin{figure}
\begin{center}
\includegraphics*[width=8cm]{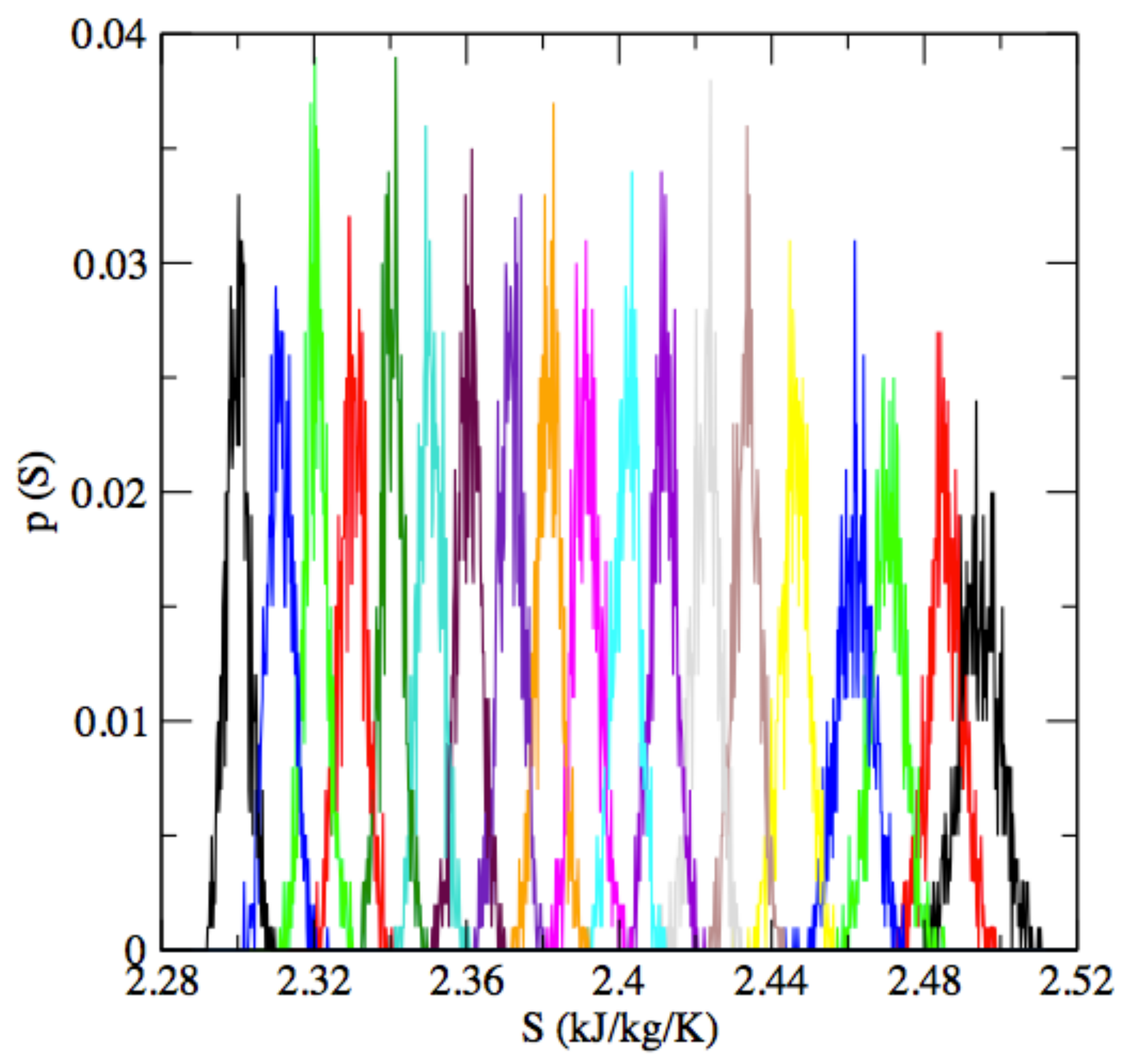}
\end{center}
\caption{$Ar$ at $T=128.76~K$. Entropy histograms $p_i(S)$ for successive sampling windows for system $2$.}
\label{Fig1}
\end{figure}

The $\mu VT-S$ simulations are initiated from a very low density vapor, obtained e.g. by placing an $Ar$ atom in the simulation cell. We then carry out a series of umbrella sampling simulations for decreasing values of the target entropy $S_{0,i}$, where $i$ is an index labeling the $i^{th}$ umbrella sampling simulation.  A reasonable estimate for the starting value for the target entropy is provided by $S_v$, the value of the vapor at coexistence. We show in Fig.~\ref{Fig1} the histograms $p_i(S)$ for the number of times a given entropy interval is visited during the $\mu VT-S$ simulations for system $2$. For each umbrella sampling simulation, comparing the relative values of $S_{max,i}$, the entropy for which the histogram $p_i(S)$ reaches its maximum, to the target entropy $S_{0,i}$, imposed during the $i^{th}$ umbrella sampling simulation, provides a great deal of insight into the nucleation process. For instance, at the beginning of the nucleation process, the system tries to overcome the free energy barrier of nucleation. This means that increasing the size of the liquid droplet (achieved here by decreasing the entropy of the system) is associated with the large cost, in terms of free energy, due to the predominant surface term. As a result, the histogram $p_i(S)$ for the entropy of the system lags behind the target value for the entropy and that $S_{max,i}>S_{0,i}$. On the other hand, when the system is past the top of the free energy barrier, the droplet has a larger probability to keep growing and we have the reverse situation with $S_{max,i}<S_{0,i}$. The same applies to very low density vapors, when the target value for the entropy is greater than the entropy of the metastable supersaturated vapor. Therefore, the entropies $S_{max,i}$ and $S_{0,i}$ only coincide when the system reaches a point corresponding to an extremum of the free energy profile, This extremum can be either a maximum (i.e. the top of the free energy barrier, when a liquid droplet of a critical size has formed) or a minimum (i.e. the metastable supersaturated vapor or the stable liquid). 

\begin{figure}
\begin{center}
\includegraphics*[width=8cm]{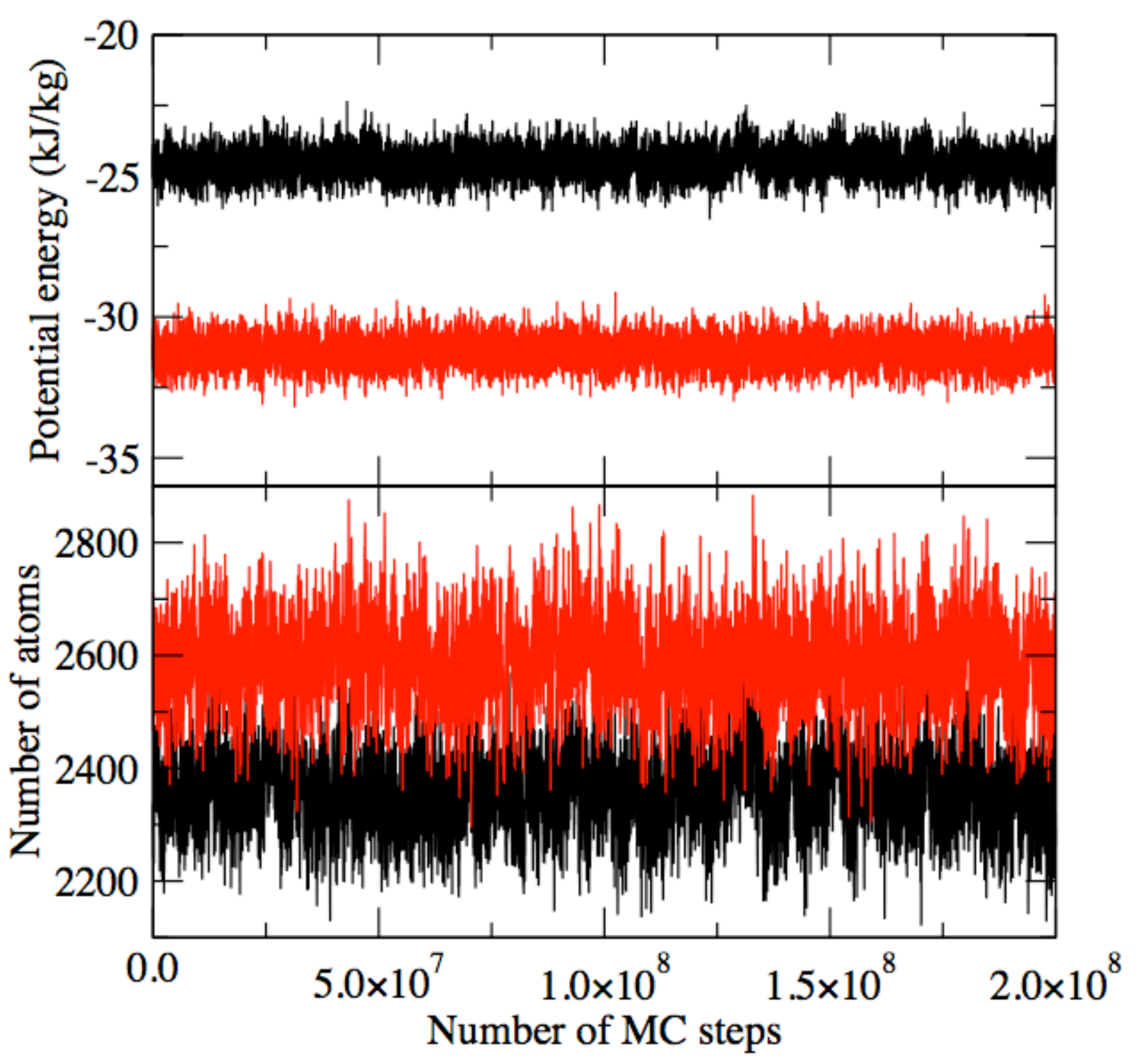}
\end{center}
\caption{$Ar$ at $T=128.76~K$. Potential energy (top) and number of atoms (bottom) for system $2$ and target entropies of $2.43$~kJ/kg/K (black) and $2.38$~kJ/kg/K (red).}
\label{Fig2}
\end{figure}

Looking at the first umbrella sampling window, starting from the right of Fig.~\ref{Fig1}, we find that $S_{max,1}$ ($2.494$~kJ/kg/K) is lower than $S_{0,1}$ ($2.5$~kJ/kg/K). This shows that, at this stage, the entropy of the system is greater than the entropy of the metastable supersaturated vapor, or, in other words, that the density of the system has not yet reached the density of the supersaturated vapor. Decreasing the target entropy allows us to identify the first value of the entropy for which $S_{max,i}=S_{0,i}$, where we have the metastable supersaturated vapor for an entropy of $2.49$~kJ/kg/K. Then, during the next umbrella sampling simulations for decreasing $S_{0,i}$, the system climbs up the free energy barrier of nucleation and we have $S_{max,i}<S_{0,i}$, until we observe again the coincidence of $S_{max,i}$ with $S_{0,i}$ for an entropy of $2.3$~kJ/kg/K. This occurs when the system reaches the top of the free energy barrier and a liquid droplet of a critical size has formed. These results show that the $\mu VT-S$ method allows us to explore the entropic pathway underlying the nucleation process and identify the extrema of the free energy profile. We also show in Fig.~\ref{Fig2} the impact of changing the target entropy on the properties of the system. Comparing the results for target entropies of $2.43$~kJ/kg/K and $2.38$~kJ/kg/K shows a decrease in the potential energy of the system by close to $30$~\% and an increase in the number of $Ar$ atoms in the system by about $10$~\%. Both results can be accounted for by the increase in the density of the system, triggered by the decrease in target entropy, which leads to an increased number of attractive interactions between $Ar$ atoms. This shows that the decrease in entropy achieved through the $\mu VT-S$ simulations results in a transition towards the liquid phase. We leave the detailed discussion of the energetics and characteristics of the liquid droplet to the 'Results' section. 

\subsubsection{Nitrogen}

The formation of a liquid droplet of $N_2$ from a supersaturated vapor is studied at $T=100~K$. As for $Ar$, we use the EWL simulation method to obtain the properties at coexistence (coex) and at three supersaturations (systems 4, 5 and 6), which are increasingly deeper inside the liquid domain of the phase diagram of $N_2$. We list in Table~\ref{muN2} the data for these points, from which we simulate the nucleation process.

\begin{table}[hbpt]
\caption{$N_2$ at $100$~K: Conditions for vapor-liquid coexistence (coex) and for three supersaturations (systems 4, 5 and 6).}
\begin{tabular}{|c|c|c|c|c|c|c|}
\hline
$ $ & $\mu$ & $\Delta \mu$  & $P$ & $P/P_{coex}$ & $S_l$ & $S_v$\\
$ $ & $(kJ/kg)$ & $(kJ/kg)$ & $(bar)$ & $ $ & $(kJ/kg/K)$ & $(kJ/kg/K)$\\
\hline
\hline
 coex & -403.97 & 0.00 & 10.12 & 1.0 & 3.374 & 4.912  \\
\hline
 system~4 & -402.16 & 1.81 & 21.84 & 2.2 & 3.352 & -  \\
\hline
 system~5 & -401.86 & 2.11 & 23.85 & 2.4 & 3.349 & -  \\
 \hline
 system~6 & -401.57 & 2.40 & 25.86 & 2.6 & 3.345 & -  \\
\hline
\hline
\end{tabular}
\label{muN2}
\end{table}

\subsubsection{Carbon dioxide}

In line with the two previous systems, we start from the conditions for vapor-liquid coexistence (coex) provided by the EWL simulations. We then define three supersaturations, for increasing $\Delta \mu$, leading to systems 7, 8 and 9. The thermodynamic data at coexistence and for the three supersaturations are given in Table~\ref{muCO2}.

\begin{table}[hbpt]
\caption{$CO_2$ at $260$~K: Thermodynamic data at the vapor-liquid coexistence (coex) and for 3 supersaturations (systems 7, 8 and 9).}
\begin{tabular}{|c|c|c|c|c|c|c|}
\hline
$ $ & $\mu$ & $\Delta \mu$  & $P$ & $P/P_{coex}$ & $S_l$ & $S_v$\\
$ $ & $(kJ/kg)$ & $(kJ/kg)$ & $(bar)$ & $ $ & $(kJ/kg/K)$ & $(kJ/kg/K)$\\
\hline
\hline
 coex & -894.03 & 0.00 & 30.22 & 1.0 & 2.962 & 3.946  \\
\hline
 system~7 & -890.84 & 3.19 & 60.28 & 2.0 & 2.944 & -  \\
\hline
 system~8 & -890.20 & 3.83 & 66.53 & 2.2 & 2.940 & -  \\
 \hline
 system~9 & -889.61 & 4.42 & 72.31 & 2.4 & 2.937 & -  \\
\hline
\hline
\end{tabular}
\label{muCO2}
\end{table}

\section{Results and Discussion}

\subsection{Argon}

\begin{figure}
\begin{center}
\includegraphics*[width=8cm]{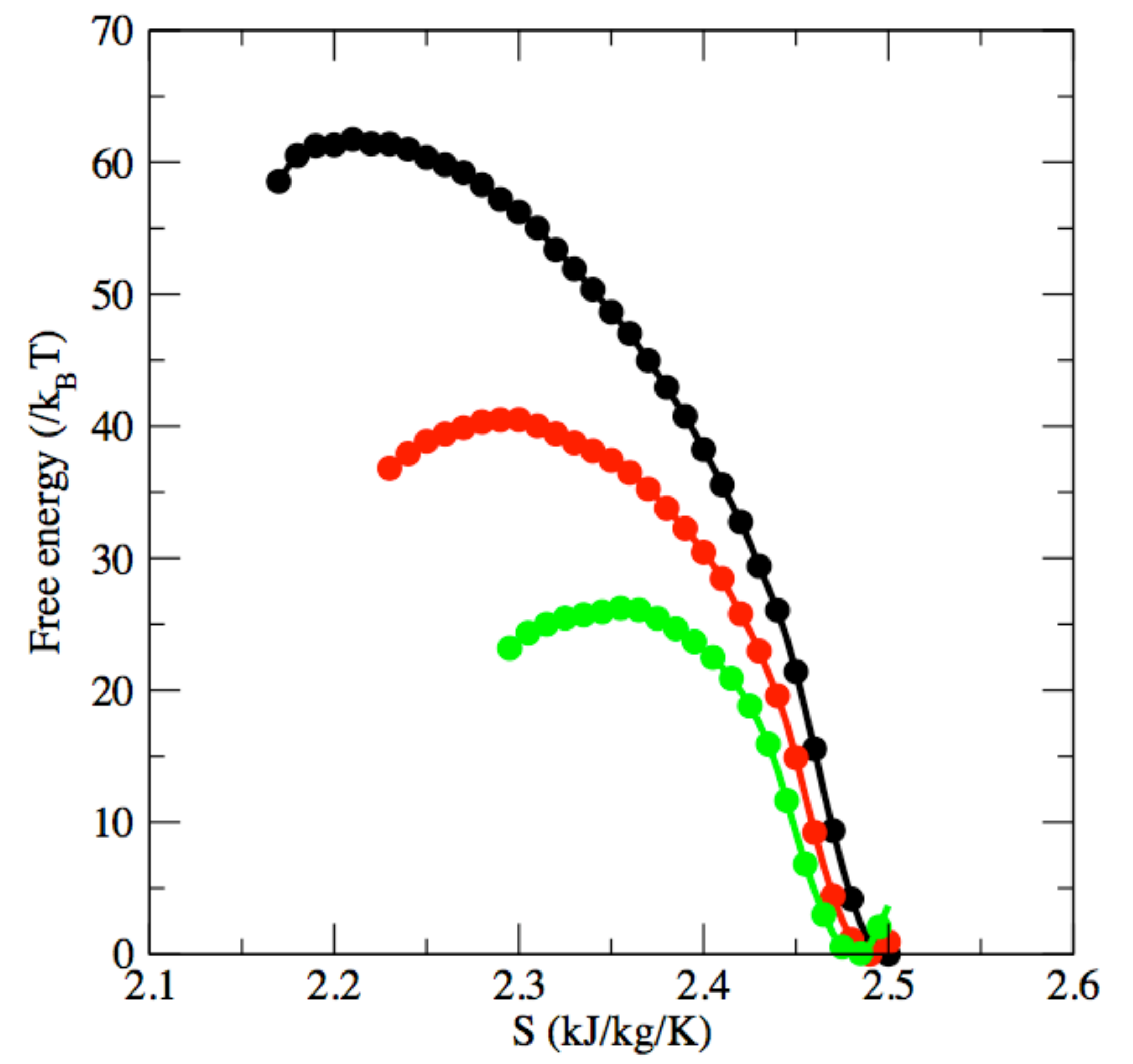}
\end{center}
\caption{$Ar$ at $T=128.76$~K: Free energy profiles of nucleation for system 1 (black), system 2 (red) and system 3 (green).}
\label{Fig3}
\end{figure}

We start by discussing the results obtained for the free energy profiles of nucleation for Argon at $T=128.76$~K. We show in Fig.~\ref{Fig3} the free energy barriers obtained from the $\mu VT-S$ simulations  for systems 1, 2 and 3 against the entropy of the system, which serves as the reaction coordinate for the nucleation process. Two main trends appear in this plot. First, the height of the free energy barrier increases as the supersaturation $\Delta \mu$ decreases. When $\Delta \mu=2.84$~kJ/kg (system 3), we obtain a free energy barrier of nucleation of $26\pm2~k_BT$, while for $\Delta \mu=2.47$~kJ/kg  (system 2), we have a free energy barrier of $41\pm3~k_BT$. Moreover, as the supersaturation further decreases to $\Delta \mu=2.04$~kJ/kg (system 1), the free energy barrier becomes $62\pm5~k_BT$. This behavior is consistent with the predictions from classical nucleation theory and with the findings from previous simulation work. Changing the supersaturation $\Delta \mu$ has a direct impact on the volume term, which is the gain, in free energy, resulting from the formation of the thermodynamically stable liquid phase from the metastable vapor phase. This volume term is proportional to $\Delta \mu$ and it therefore follows that for a small $\Delta \mu$ (system 1), the top of the free energy barrier is reached later, for a larger liquid droplet, leading to a higher free energy barrier. The values obtained in this work for the free energy barriers of nucleation are also in good agreement with those predicted by the classical nucleation theory~\cite{mcgraw1996scaling,mcgraw1997interfacial} and with those found for the Lennard-Jones system~\cite{ten1998computer}. For instance, CNT leads to the following free energy of nucleation~\cite{mcgraw1996scaling,mcgraw1997interfacial} $W=16\pi \gamma^3/(3\rho_l^2\Delta \mu^2)$, where $\gamma$ is the surface tension for a flat interface and $\rho_l$ is the density of the (bulk) liquid. Applying this to e.g. System 1 with $\gamma=3.535 \times 10^{-10} kJ/cm^2$ (determined from the EWL simulations~\cite{PartI}) and $\rho_l=1.0882~g/cm^3$ leads to a free energy barrier of $70~k_BT$, in reasonably good agreement with the barrier found here of $62\pm5~k_BT$. Our results are also consistent with ten Wolde and Frenkel~\cite{ten1998computer} who reported free energy barriers of nucleation $5-15~k_BT$ lower than the CNT predictions for the Lennard-Jones system. The second main trend is the following. The entropic pathway spanned during nucleation covers a range of entropy that depends on the supersaturation. More specifically, for a large supersaturation, the entropy range becomes narrower. This is due to the fact that for a low supersaturation (i.e. not too deep into the liquid domain of the phase diagram), the metastable supersaturated vapor is less dense. This means that the starting point for the nucleation process is associated with a larger entropy. Furthermore, at low supersaturation, the size of the critical nucleus is larger. This implies that, when the supersaturation is low, the entropy of the system at the top of the free energy barrier is smaller. Both factors account for the extended range of entropy spanned during the nucleation process at low supersaturation. We finally add that the range of entropy spanned during nucleation is, for all supersaturations, well within the interval defined by the entropies of the vapor ($2.701$~kJ/kg/K) and liquid ($1.856$~kJ/kg/K) at coexistence. This is the expected behavior, since the metastable supersaturated vapors are more dense, with a smaller entropy, than the vapor at coexistence, and since systems containing a critical liquid droplet are considerably less dense, with a larger entropy, than a uniform liquid phase like the liquid at coexistence.

\begin{figure}
\begin{center}
\includegraphics*[width=7cm]{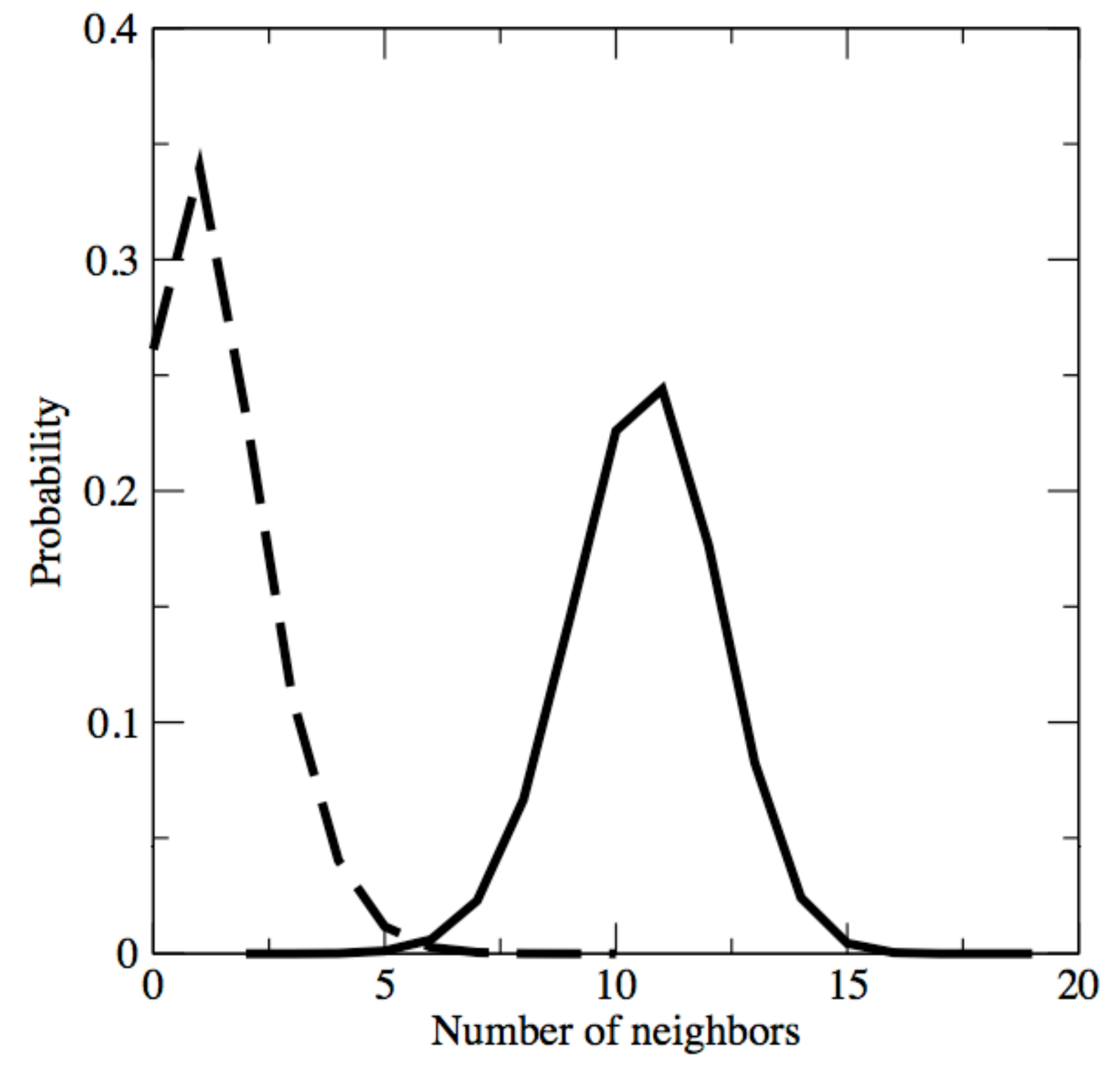}(a)
\includegraphics*[width=7cm]{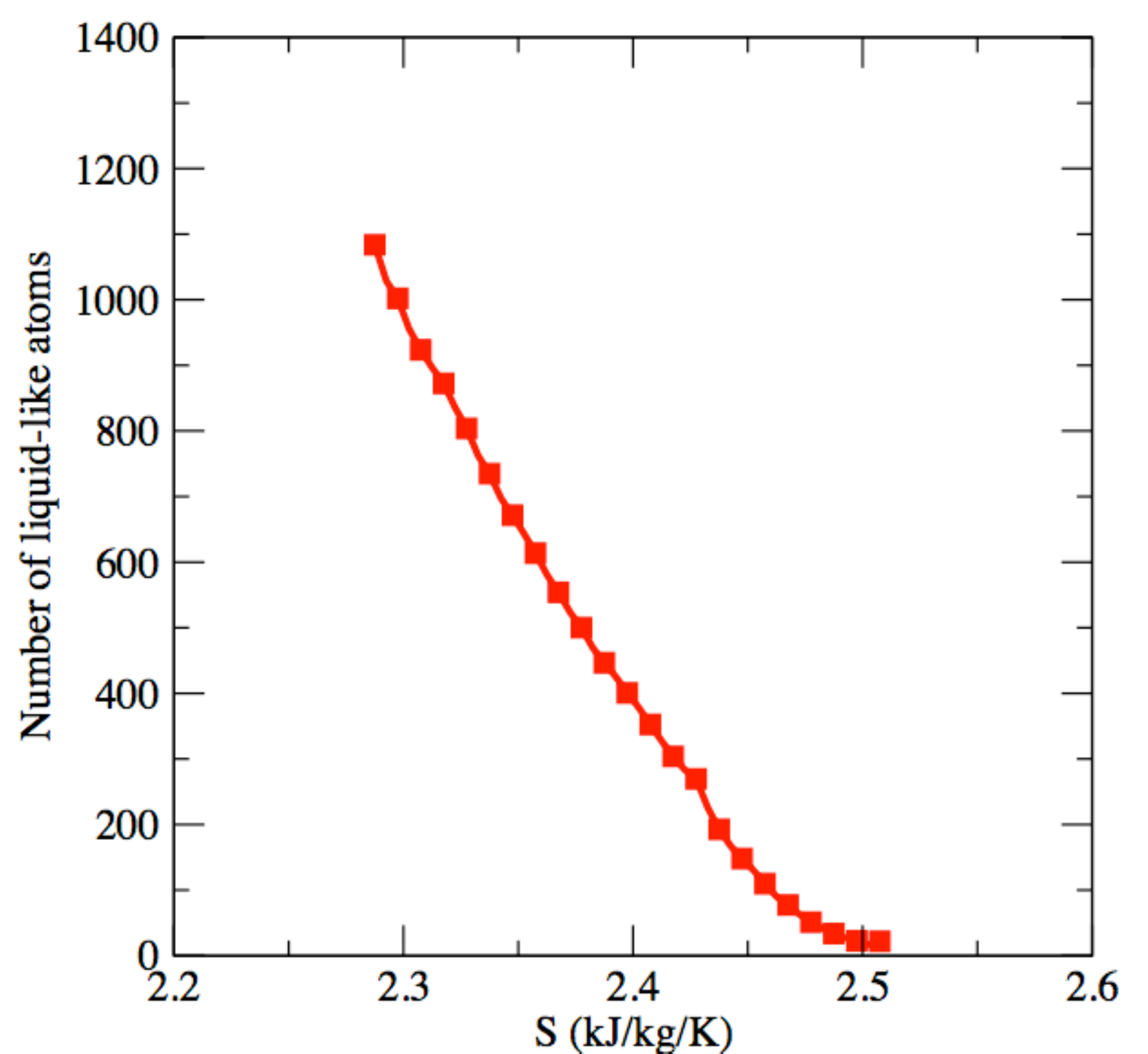}(b)
\end{center}
\caption{$Ar$ at $T=128.76$~K: (a) Distribution for the number of neighbors in the vapor (dashed line) and in the liquid (solid line) and (b) Correspondence between the entropy and the number of liquid-like atoms along the nucleation (system 2).}
\label{Fig4}
\end{figure}

\begin{figure}
\begin{center}
\includegraphics*[width=6cm]{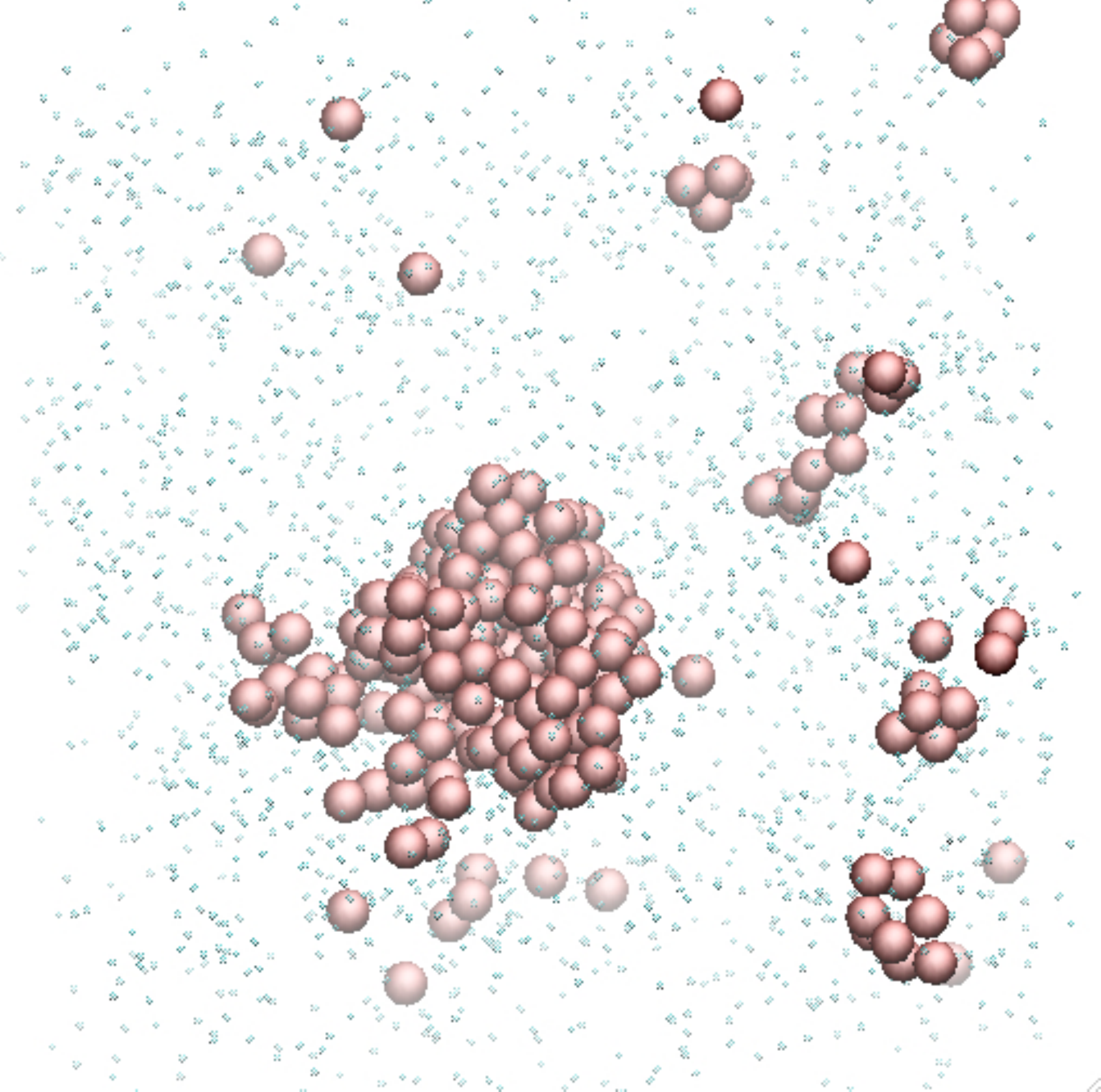}(a)
\includegraphics*[width=6cm]{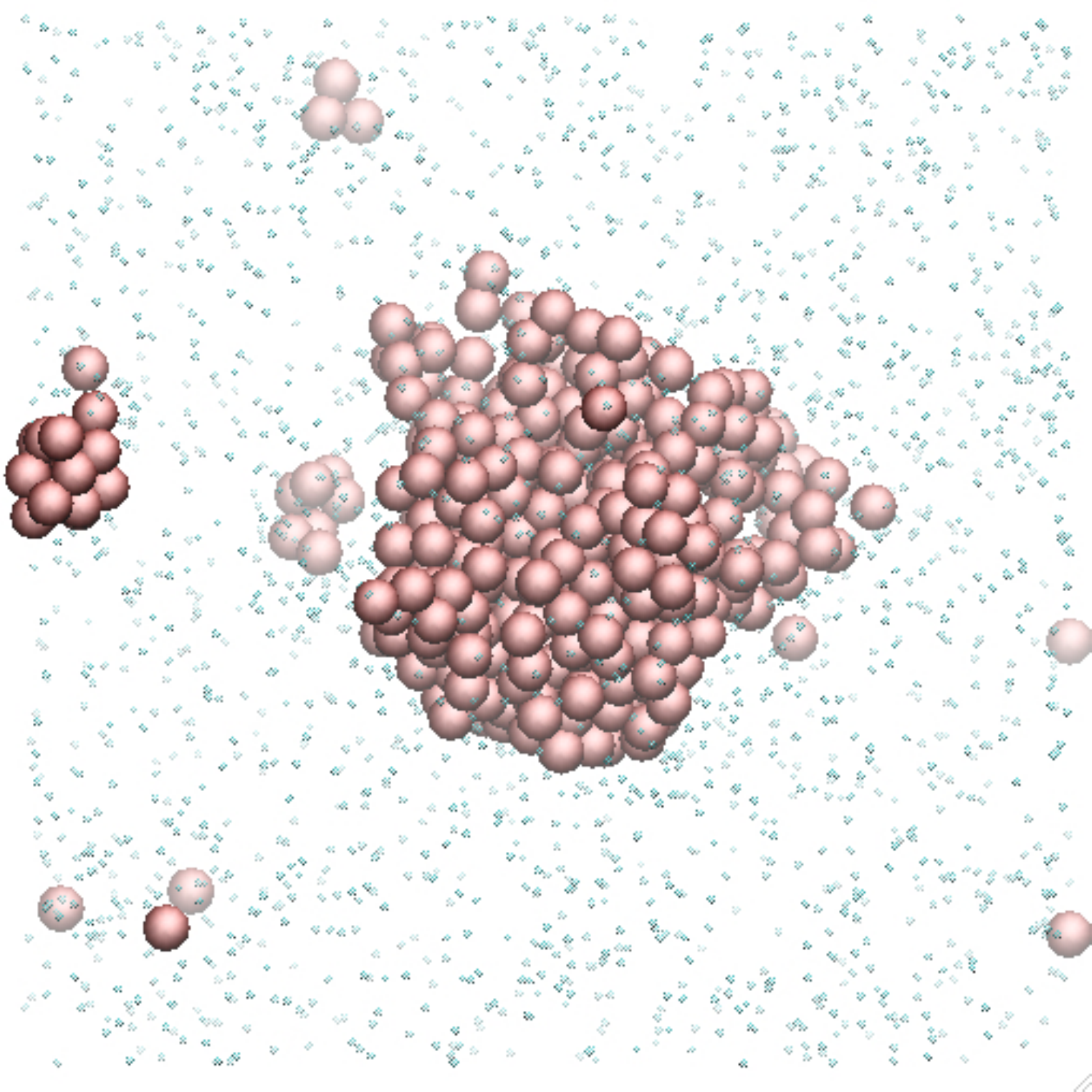}(b)
\includegraphics*[width=6cm]{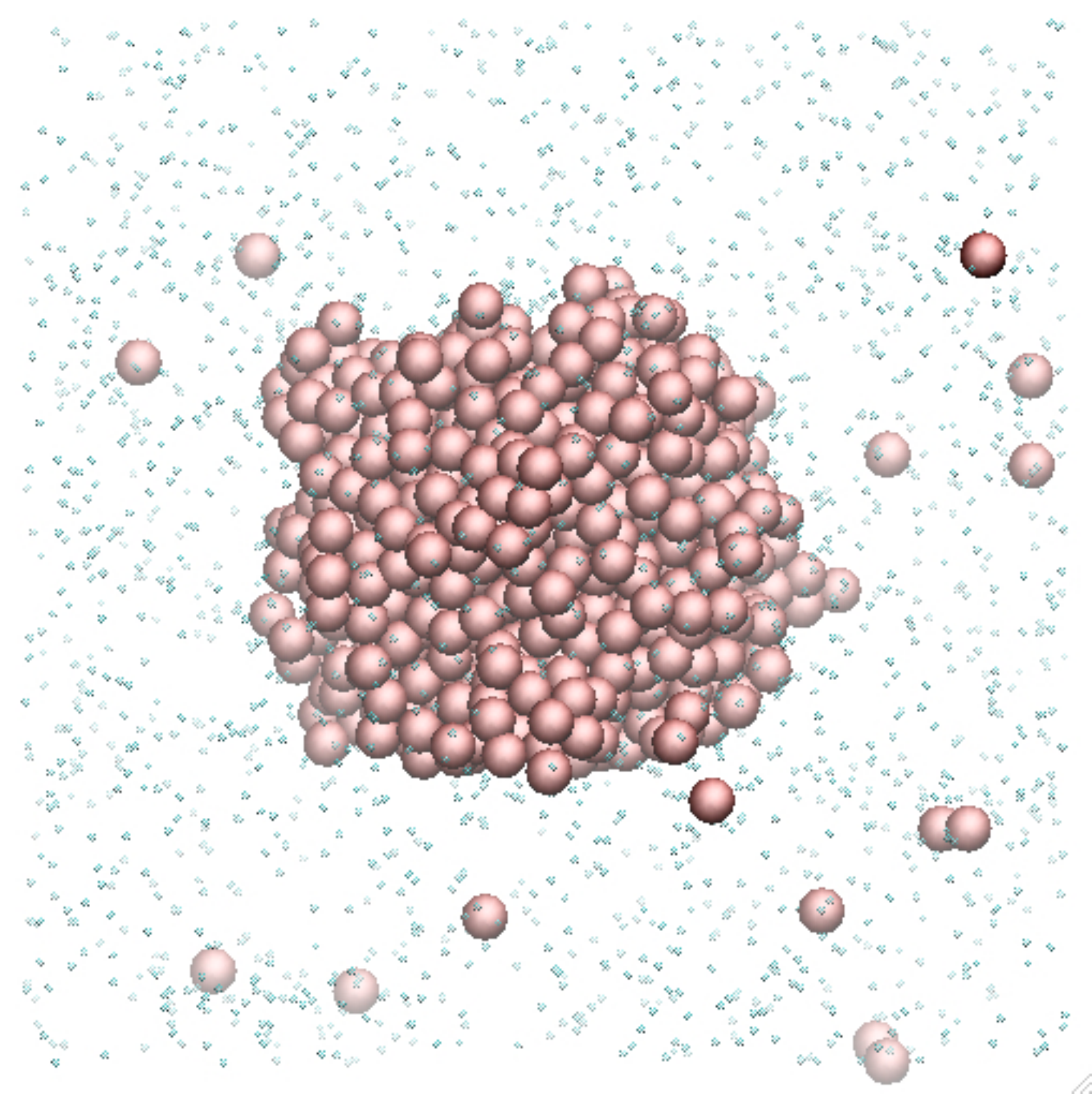}(c)
\end{center}
\caption{$Ar$ at $T=128.76$~K: Snapshots for system 1 during the nucleation process for decreasing values of the entropy: (a) $S=2.43$~kJ/kg/K, (b) $S=2.38$~kJ/kg/K, (c) $S=2.31$~kJ/kg/K }
\label{Fig5}
\end{figure}

The plots in Fig.~\ref{Fig3} also allow us to characterize the state of the system at the top of the free energy barrier of nucleation in terms of the critical entropy associated with the formation of a critical nucleus. This critical entropy $S_c$ is shown to depend on supersaturation, with the largest value for $S_c$ being obtained for the lowest supersaturation $S_c=2.21\pm0.02$~kJ/kg (system 1) and the smallest value $S_c=2.35\pm0.01$~kJ/kg for the highest supersaturation (system 3). To analyze further the characteristics of the liquid droplet, we examine the relation between the entropy and the size of the liquid droplet. Starting from the approach developed by ten Wolde and Frenkel~\cite{ten1998computer}, we first determine the distribution for the number of neighbors in the vapor and in the liquid. Here, neighboring particles are defined as particles that are less than $1.6 \sigma$ apart. This distance of $1.6 \sigma$ corresponds to the first minimum of the radial pair distribution function of the liquid. Fig~\ref{Fig4}(a) shows the distribution so obtained for the vapor and for the liquid phases. The two distributions show little overlap, with the vast majority of the liquid atoms having at least $6$ neighbors, while atoms in the vapor phase very rarely have more than $5$ neighbors. This allows us to define liquid-like atoms as having at least $6$ neighbors and vapor-like atoms having fewer than $6$ neighbors and to follow the variations of the number of liquid-like atoms during nucleation. Applying this analysis during the $\mu VT-S$ simulations leads us to draw a correspondence between the value of the entropy and the number of liquid-like atoms  of the system as nucleation proceeds. We present in Fig.~\ref{Fig4}(b) the result for this correspondence for system 2. The number of liquid-like atoms increases steadily with entropy as the nucleation process advances, implying that the size of the liquid droplet steadily increases as the entropy of the system decreases. This can best be seen by examining snapshots of the configurations of the system during the nucleation process. Fig.~\ref{Fig5} shows the formation and development of the liquid droplet along the entropic pathway. As can be seen on the snapshots, the droplet size steadily increases with entropy and reaches a critical size of $1715\pm50$ atoms for system 1. The snapshots for the other 2 supersaturations reveal a similar behavior, and critical sizes of $980\pm25$ and $649\pm15$ atoms are found for the two other supersaturations systems 2 and 3, respectively. We obtain the expected correlation between the height of the free energy barrier of nucleation and the size of the critical nucleus, as the size of the critical droplet decreases with the free energy of nucleation as we go from system 1 to system 2 and finally to system 3.

\subsection{Nitrogen}

\begin{figure}
\begin{center}
\includegraphics*[width=8cm]{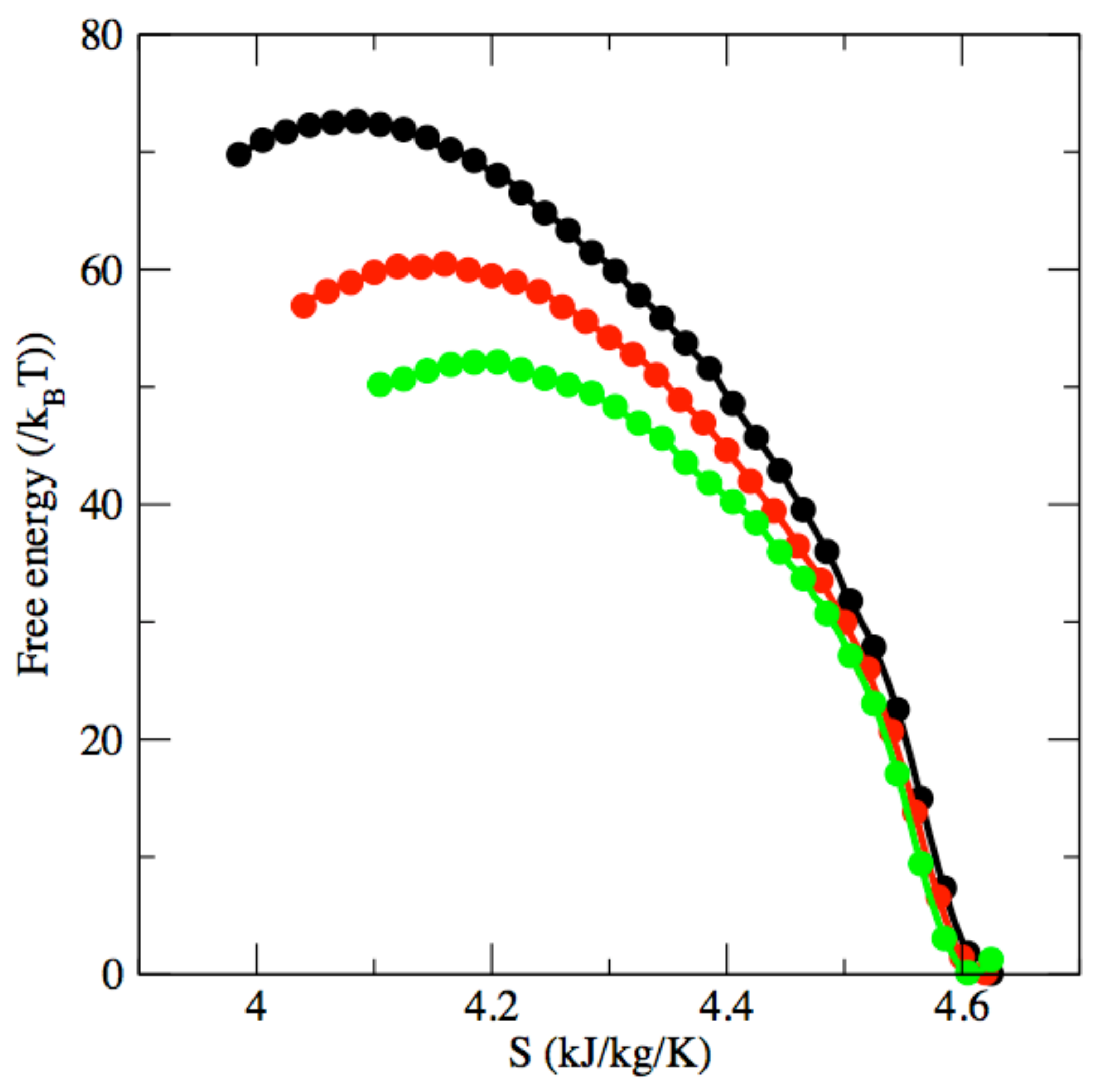}
\end{center}
\caption{$N_2$ at $T=100$~K: Free energy profiles of nucleation along $S$ for system 4 (black), system 5 (red) and system 6 (green).}
\label{Fig6}
\end{figure}

We now turn to the assessment of the versatility of the method and test it on a molecular fluid. We apply the $\mu VT-S$ approach to study the nucleation of a liquid droplet of $N_2$ at $100$~K. We follow the same procedure as for $Ar$ and perform a series of umbrella sampling simulations spanning the entropic pathway underlying the formation of the droplet. We present in Fig.~\ref{Fig6} the free energy profiles of nucleation of $N_2$ at $T=100$ for three supersaturations (systems 4, 5 and 6). The results show features that are qualitatively similar to those found for $Ar$. In terms of entropy range, the entropies sampled during the process lie within the wider interval defined by the entropy of the liquid and of the vapor at coexistence (from $S_{liq}=3.374$~kJ/kg/K to $S_{vap}=4.912$~kJ/kg/K). For instance, for the lowest supersaturation (system 4), the nucleation starts from a supersaturated vapor with $S=4.6$~kJ/kg/K (lower than the entropy of the coexisting vapor, that has a lower density) and ends with configurations containing a critical nucleus at $S=4.08$~kJ/kg/K (higher than the entropy of the liquid at coexistence, which has a much a higher density). The same reasoning applied to systems 5 and 6, for which the nucleation process is associated to an even narrower entropy range due to the higher supersaturations involved for these systems. In particular, the entropy $S_c$ for which a liquid droplet of a critical size has formed is shown to increase as the supersaturation decreases, starting from $S_c=4.08\pm 0.02$~kJ/kg/k (system 4) and increasing to $S_c=4.16\pm 0.02$~kJ/kg/K (system 5) and finally to $S_c=4.20\pm 0.02$~kJ/kg/K (system 6) for the highest supersaturation. The free energy barriers of nucleation for $N_2$ are also shown to decrease gradually as the supersaturation is increased. The free energy barrier of nucleation reaches $73\pm5~k_BT$ for system 4 (lowest supersaturation), $60\pm 4~k_B T$ for system 5 and $52 \pm 4~k_B T$ for system 6 (highest supersaturation).

\begin{figure}
\begin{center}
\includegraphics*[width=7cm]{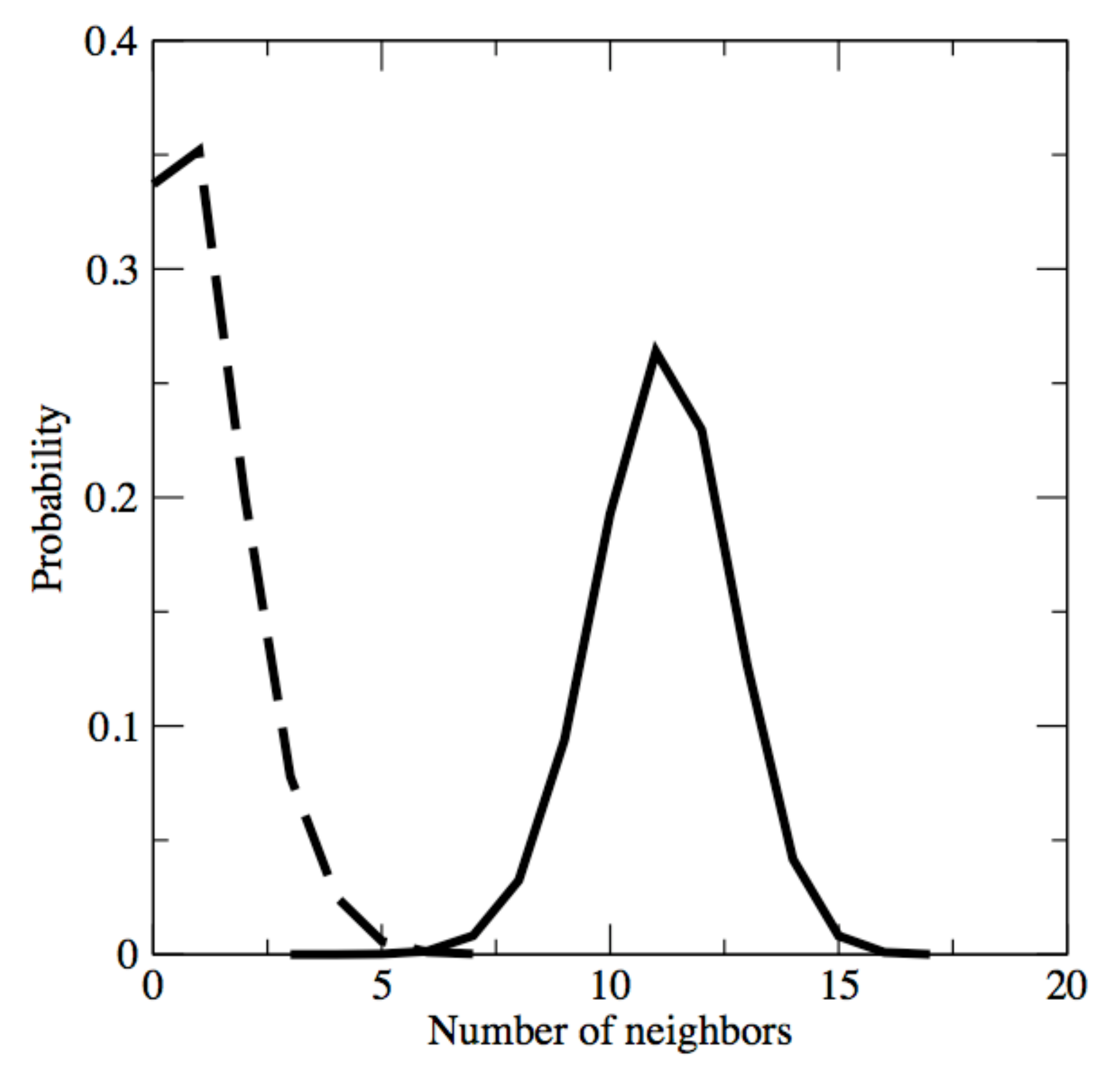}(a)
\includegraphics*[width=7cm]{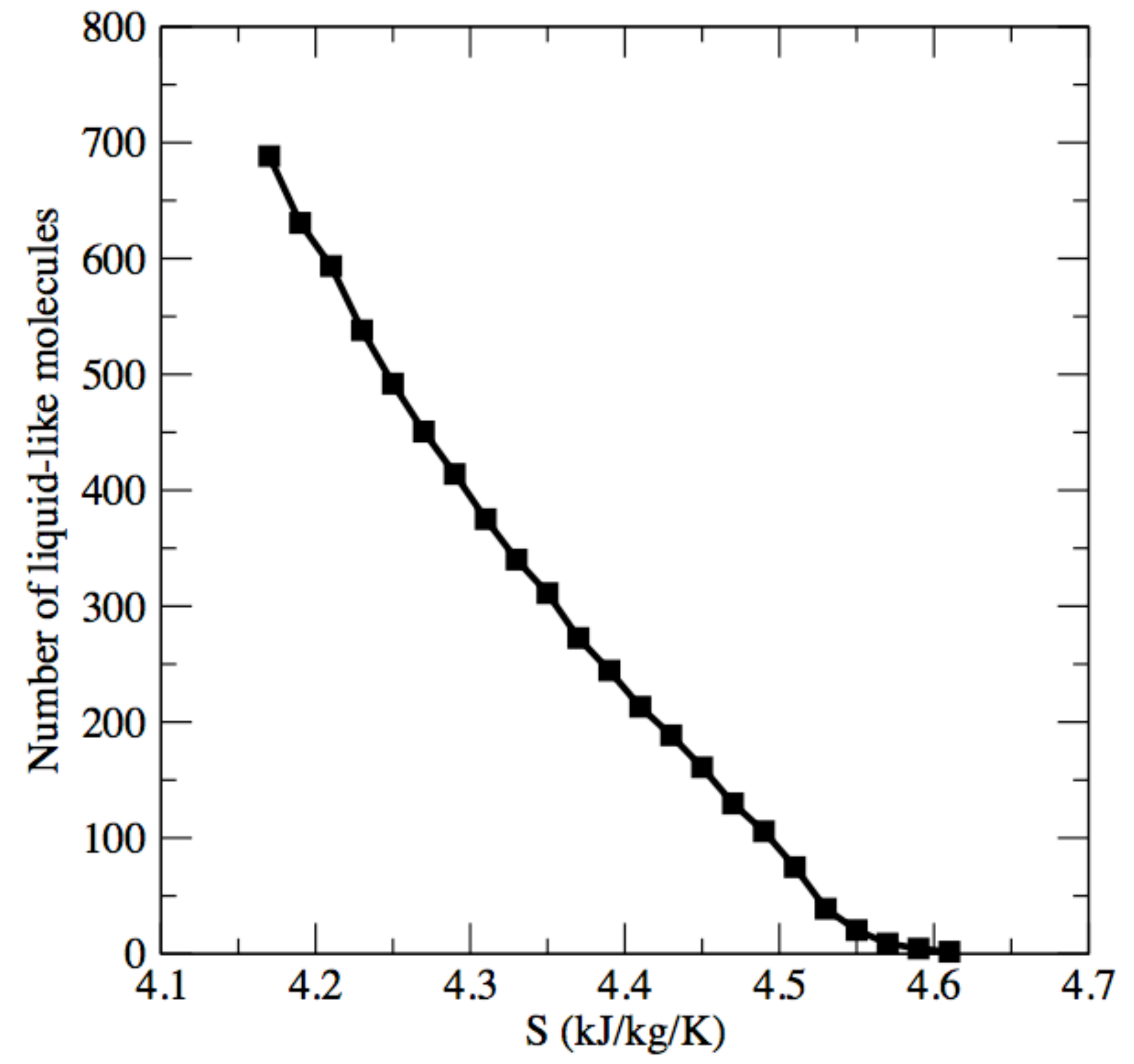}(b)
\end{center}
\caption{$N_2$ at $T=100$~K: (a) Distribution for the number of neighbors in the vapor (dashed line) and in the liquid (solid line) and (b) Correspondence between the entropy and the number of liquid-like molecules during the nucleation process (system 6).}
\label{Fig7}
\end{figure}

\begin{figure}
\begin{center}
\includegraphics*[width=6cm]{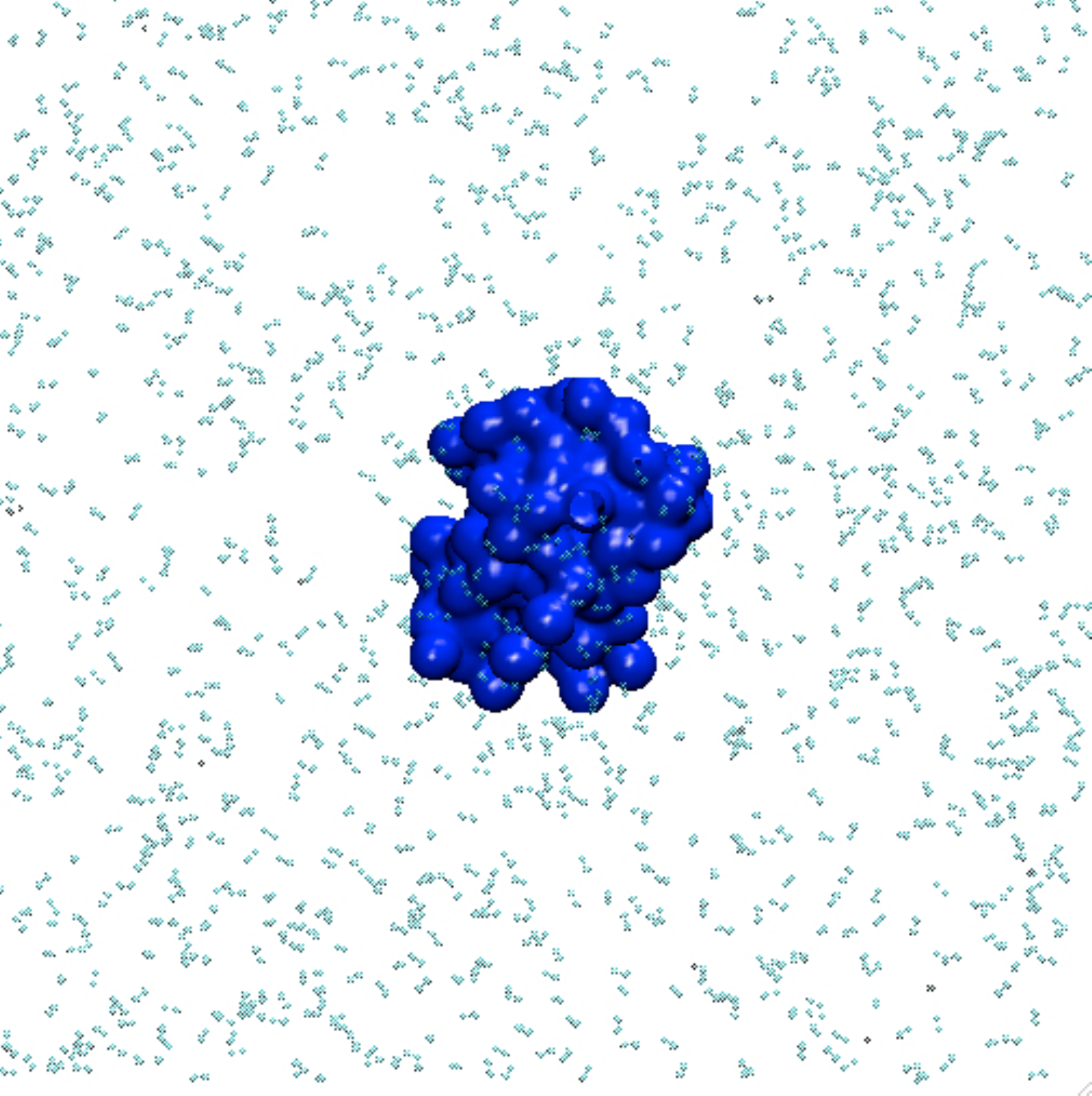}(a)
\includegraphics*[width=6cm]{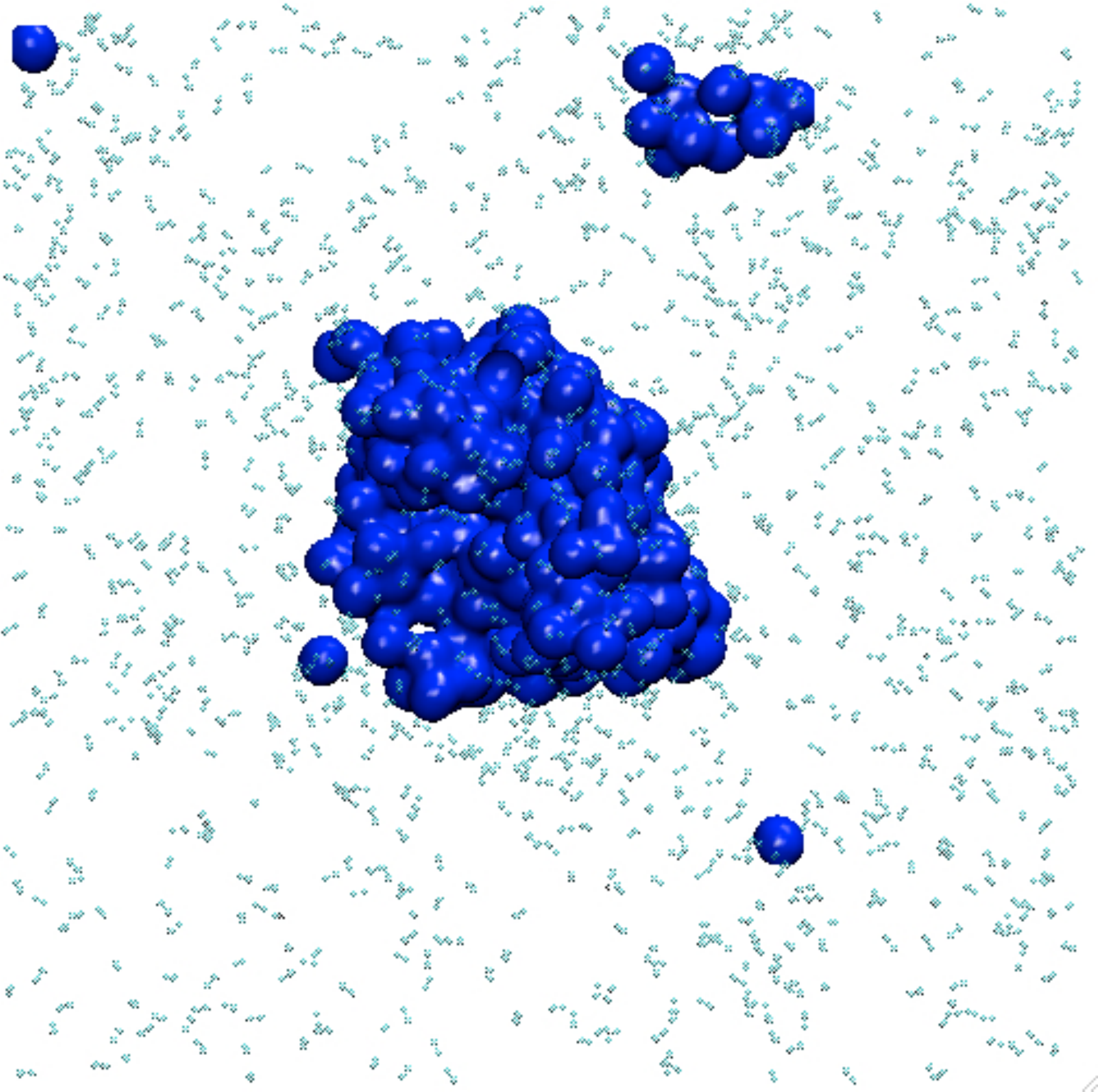}(b)
\includegraphics*[width=6cm]{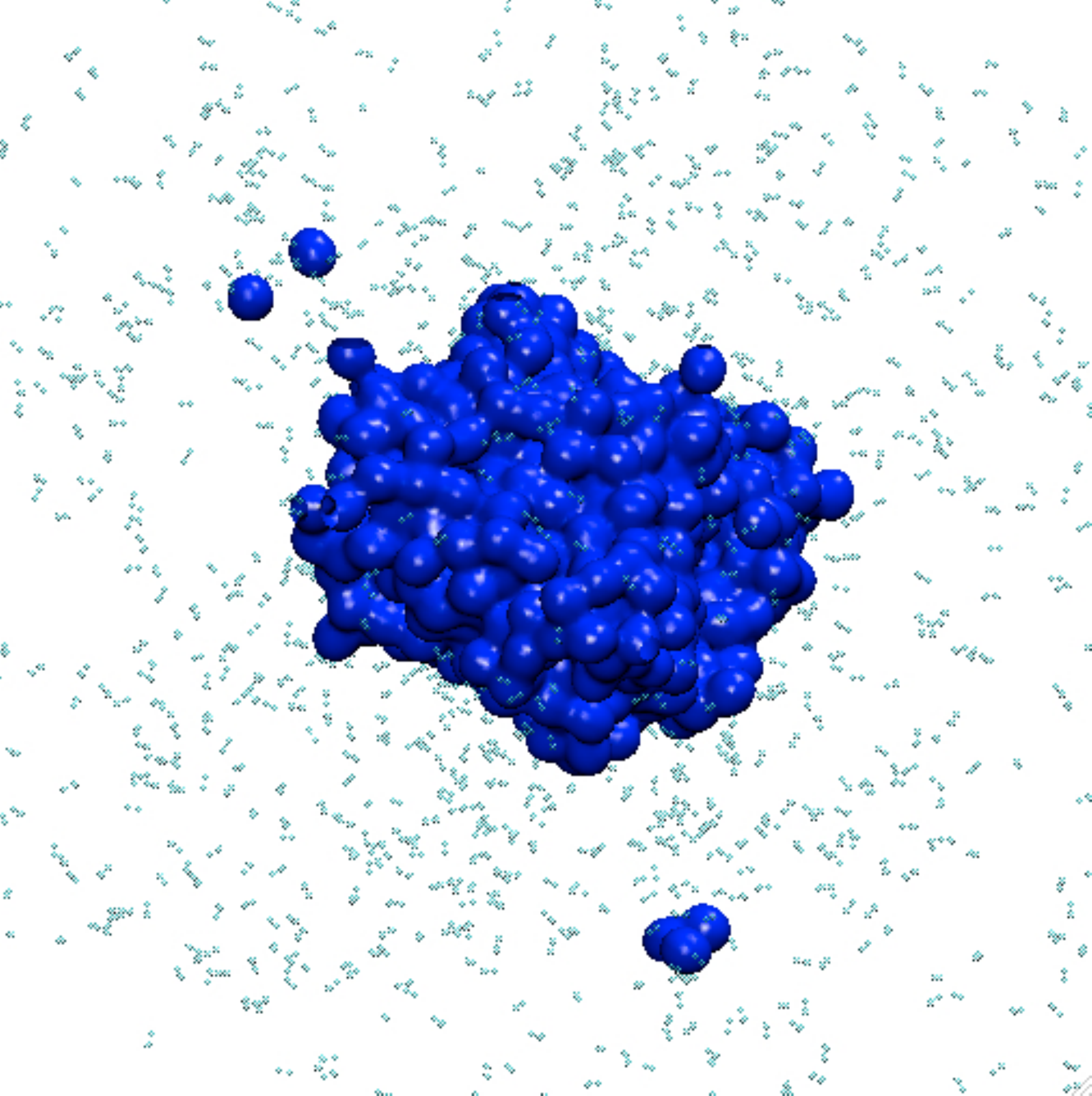}(c)
\end{center}
\caption{$N_2$ at $T=100$~K: Snapshots of system 6 during the nucleation process for decreasing values of the entropy: (a) $S=4.5$~kJ/kg/K, (b) $S=4.36$~kJ/kg/K, (c) $S=4.2$~kJ/kg/K }
\label{Fig8}
\end{figure}

We further refine our analysis by determining the size of the droplet during the nucleation process. For this purpose, we adopt a similar procedure to that followed for Argon. We compute the distributions for the number of neighbors for $N_2$ molecules in the supersaturated vapor and in the liquid, using the first minimum of the pair distribution function (based on the distance between the centers-of-mass of 2 $N_2$ molecules) at $5.7$~\AA~as a cutoff distance used to define neighbors. The results plotted in Fig.~\ref{Fig7}(a) show that these distributions allow us to identify vapor-like molecules (with less than $6$ neighbors) and liquid-like molecules (with $6$ or more neighbors). The variation of the number of liquid-like molecules in the system as a function of entropy during the nucleation process is shown in Fig.~\ref{Fig7}(b). As with $Ar$,the number of liquid-like molecules exhibits a steady increase with entropy, corresponding to the increase in the size of the droplet as nucleation proceeds. An examination of the snapshots (see Fig.~\ref{Fig8}) obtained during the $\mu VT-S$ simulations sheds light on the formation of the liquid droplet as the entropy of the system decreases. This allows us to identify the size of the critical droplet, which contains $1006\pm30$ $N_2$ molecules for system 4, $735\pm25$ molecules for system 5 and $638\pm20$ molecules for system 6. We observe the expected correlation between the size of the critical droplet and the height of the free energy barrier of nucleation, with the lowest supersaturation of system 4 leading to the largest free energy of nucleation and to the biggest critical droplet, that are reached for the lowest critical entropy. This set of results show that $S$ can be used as the reaction coordinate for the nucleation process in a molecular fluid, and that no adjustment or redefinition of the RC is necessary.

\subsection{Carbon dioxide}

\begin{figure}
\begin{center}
\includegraphics*[width=8cm]{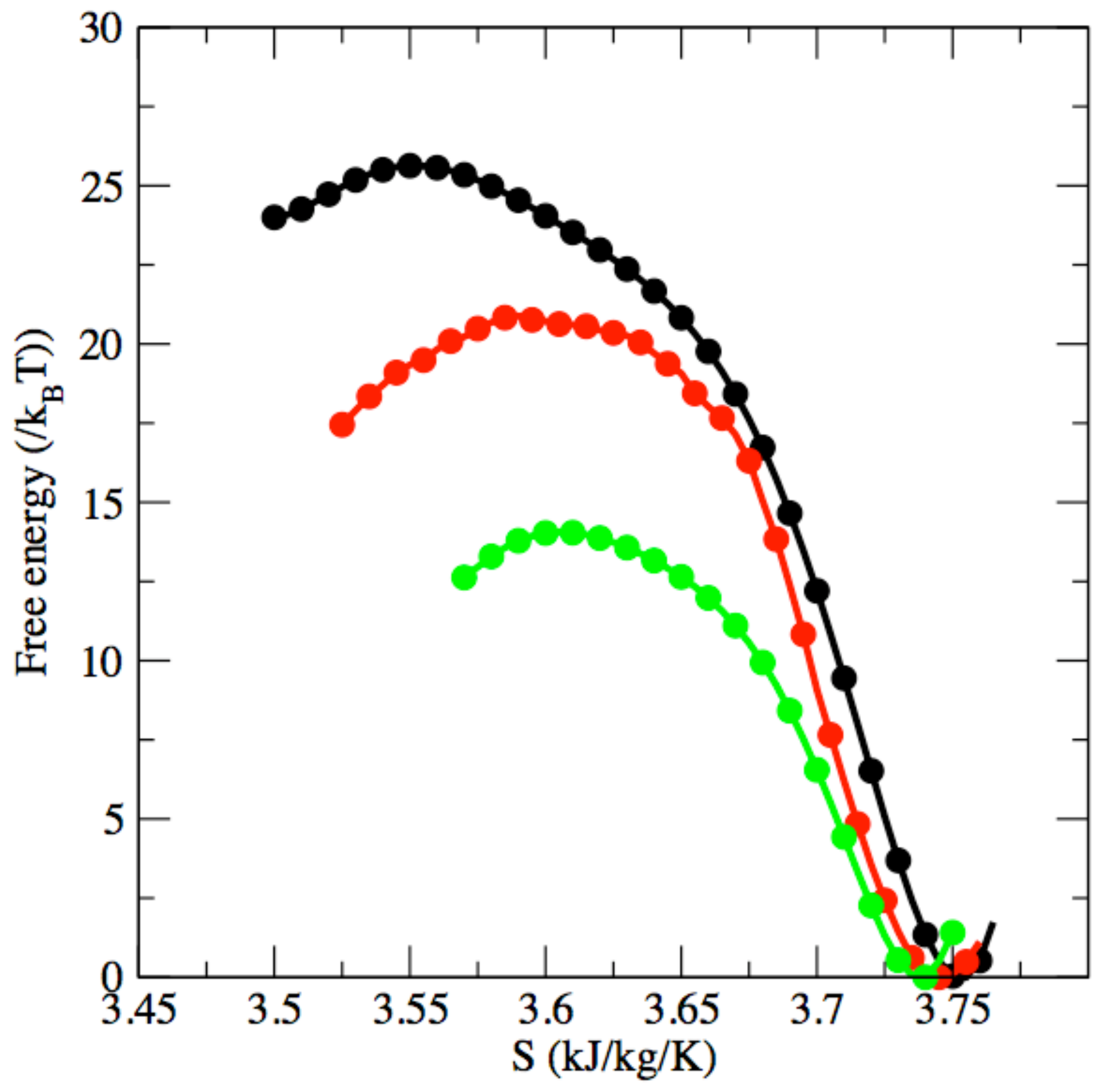}
\end{center}
\caption{$CO_2$ at $T=260$~K: Free energy profiles of nucleation along $S$ for system 7 (black), system 8 (red) and system 9 (green).}
\label{Fig9}
\end{figure}

We now apply the $\mu VT-S$ method to a second molecular fluid and study the nucleation of a liquid droplet in $CO_2$ at $260~K$. $CO_2$ is similar to $N_2$ in many respects, having a similar shape and exhibiting similar types of intermolecular interactions (with a quadrupole moment for $CO_2$ more than 6 times greater than for $N_2$). Carrying out the $\mu VT-S$ simulations for the formation of the liquid droplet at $260$~K allows us to draw a comparison with the results on $N_2$. This is because the conditions of temperature for $N_2$ and $CO_2$ can be considered as similar, if one considers reduced temperatures, with respect to the critical temperature, for each compounds ($0.85$ for $CO_2$ and $0.83$ for $N_2$). As for $Ar$ and $N_2$, we perform a series of $\mu VT-S$ simulations to sample the entropic pathway leading towards the formation of a liquid droplet of $CO_2$ of a critical size for 3 supersaturations. We show in Fig.~\ref{Fig9} the free energy profiles of nucleation so obtained. Similarly to the nucleation of liquid droplet in $N_2$, the profiles show that decreasing the entropy of the system during the $\mu VT-S$ simulations allows the system to overcome the free energy barrier and that $S$ can be reliably used to drive the nucleation process in the molecular fluid of $CO_2$.  

\begin{figure}
\begin{center}
\includegraphics*[width=7cm]{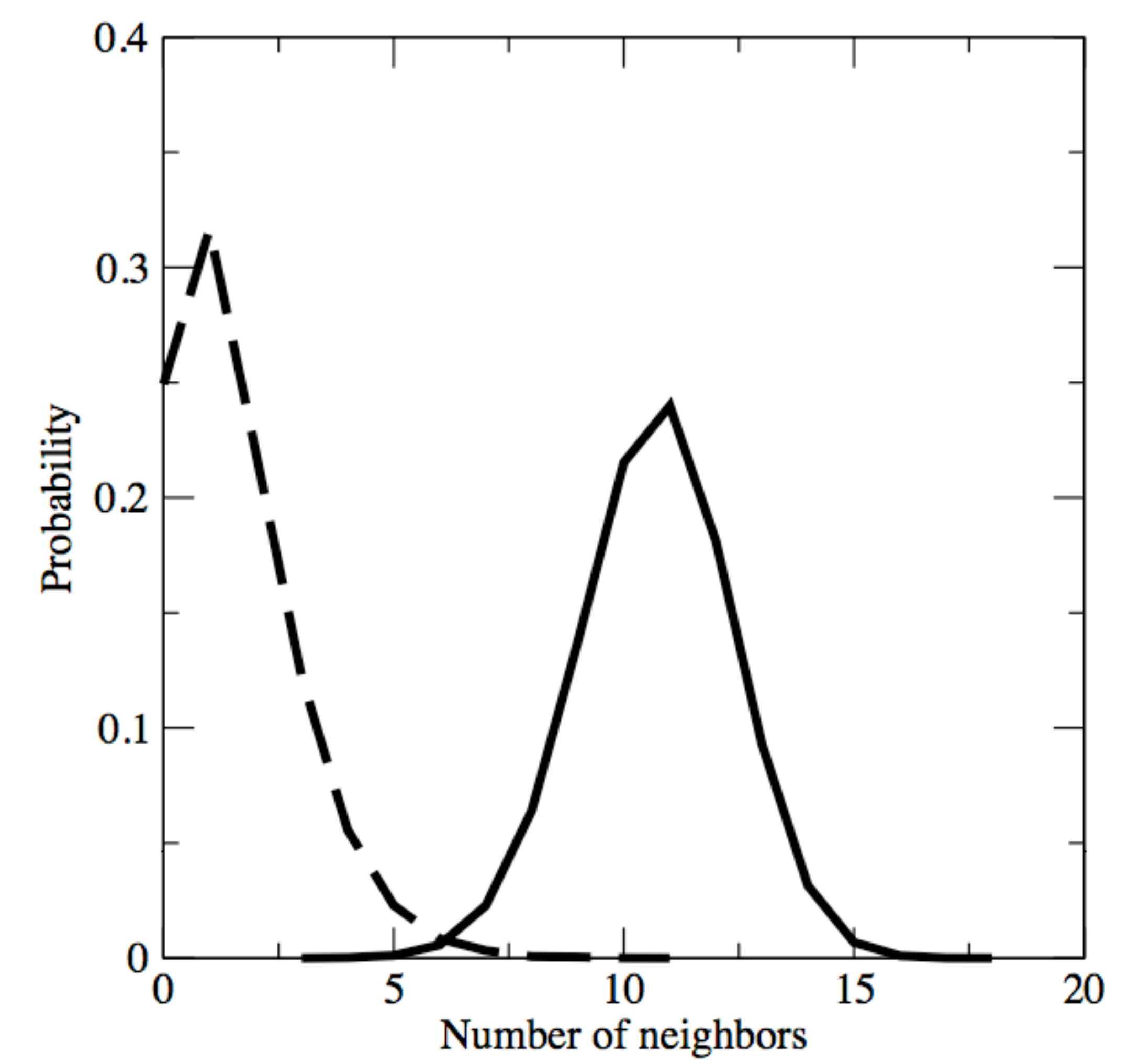}(a)
\includegraphics*[width=7cm]{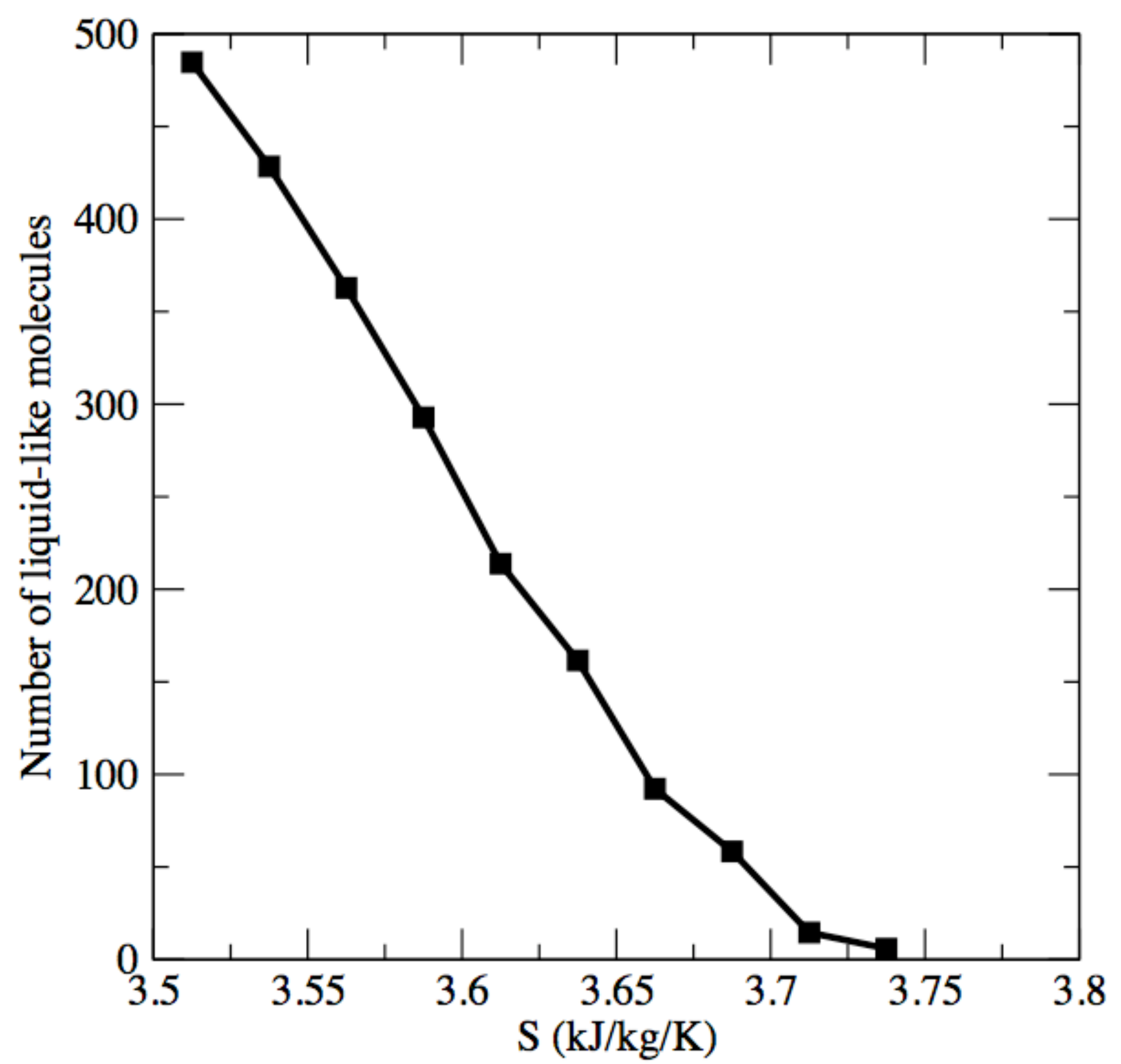}(b)
\end{center}
\caption{$CO_2$ at $T=260$~K: (a) Distribution for the number of neighbors in the vapor (dashed line) and in the liquid (solid line) and (b) Correspondence between the entropy and the number of liquid-like molecules during nucleation (system 7).}
\label{Fig10}
\end{figure}

The range of entropies sampled along the entropic pathway is shown to depend on the extent of supersaturation. As for $Ar$ and $N_2$, the entropy for the starting point, the metastable supersaturated vapor, decreases at high supersaturations as the supersaturated vapor becomes more dense. The entropy for which the top of the free energy barrier is reached also decreases as the supersaturation becomes high, as a result of the smaller size of the critical droplet. Both factors account for the reduced range of entropies sampled for system 9 (high supersaturation) when compared to system 7 (low supersaturation). This leads to the following values for the entropies $S_c$ corresponding to systems containing a liquid droplet of a critical size, with $S_c=3.55\pm 0.02$~kJ/kg/k (system 7) and decreasing to $S_c=3.59\pm 0.02$~kJ/kg/K (system 8) and finally to $S_c=3.61\pm 0.02$~kJ/kg/K (system 9) for the highest supersaturation. This increase in $S_c$ at high supersaturations is correlated with a decrease in the height of the free energy barrier of nucleation which reaches $26\pm3~k_BT$ for system 7 (lowest supersaturation), $21\pm 4~k_B T$ for system 8 and $14 \pm 2~k_B T$ for system 9 (highest supersaturation). Since the reduced temperatures and the supersaturations for the nucleations of $N_2$ and $CO_2$ are similar, we attribute the much smaller free energy barriers of nucleation, when compared to those found for $N_2$, to the stronger intermolecular interactions in $CO_2$.

\begin{figure}
\begin{center}
\includegraphics*[width=6cm]{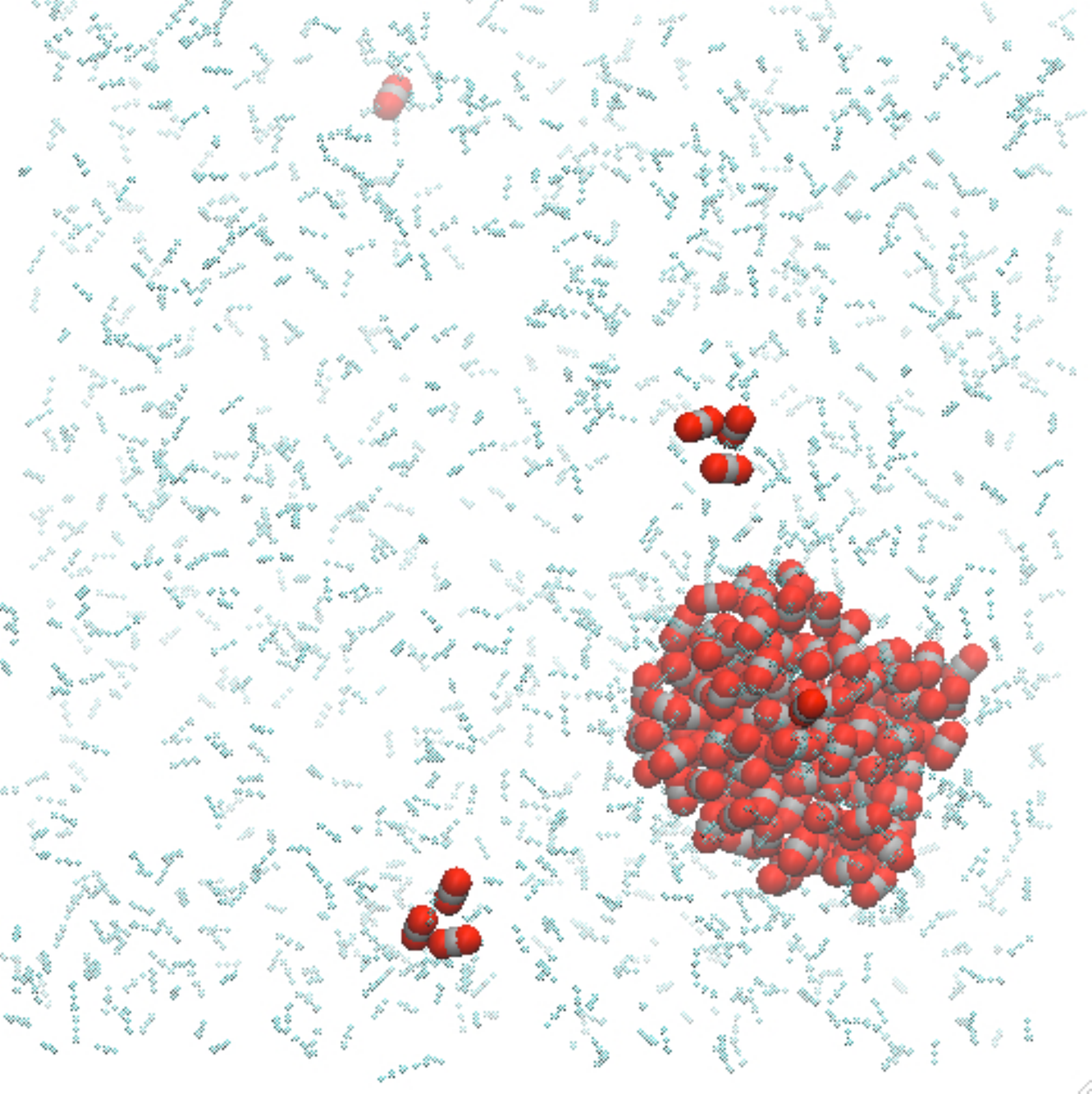}(a)
\includegraphics*[width=6cm]{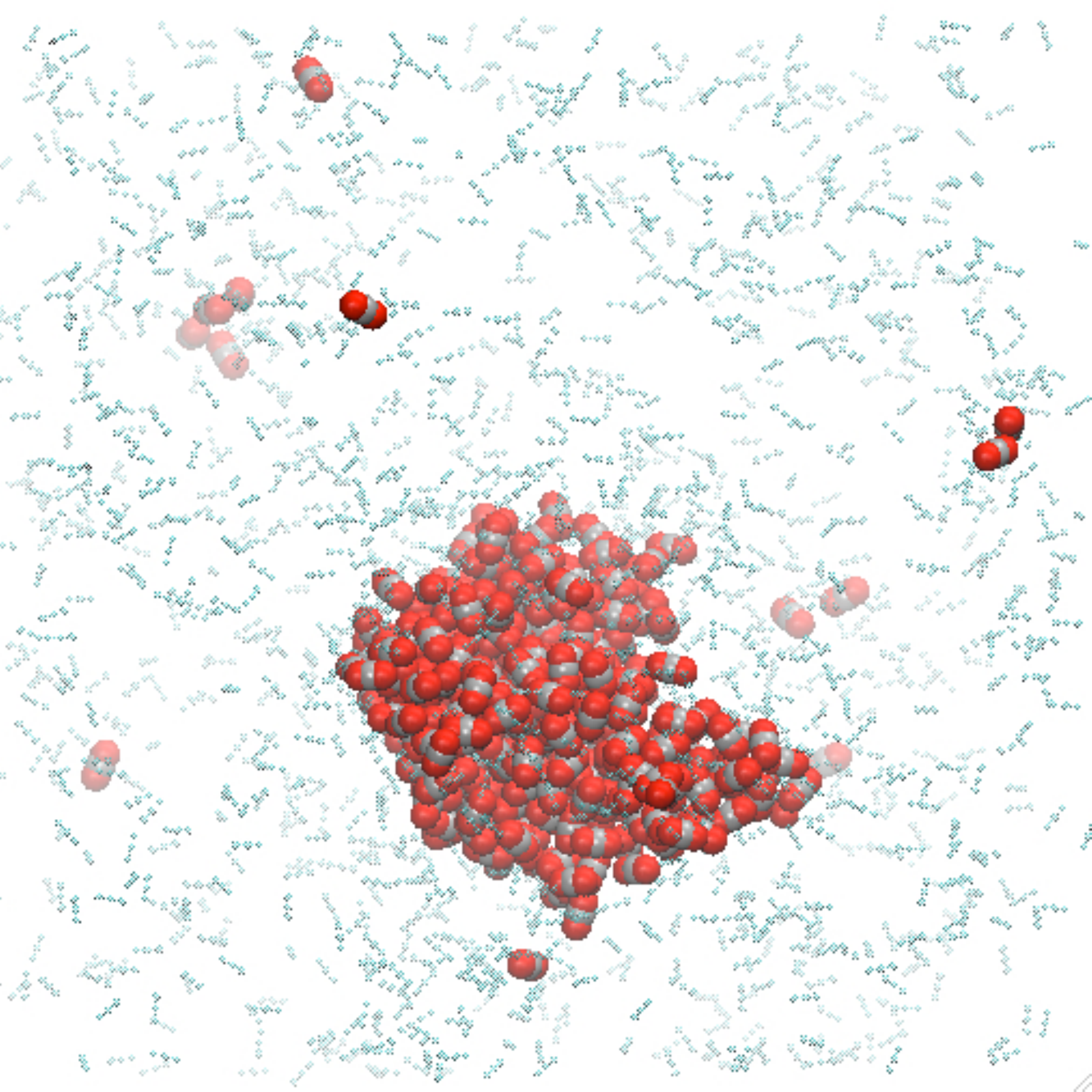}(b)
\includegraphics*[width=6cm]{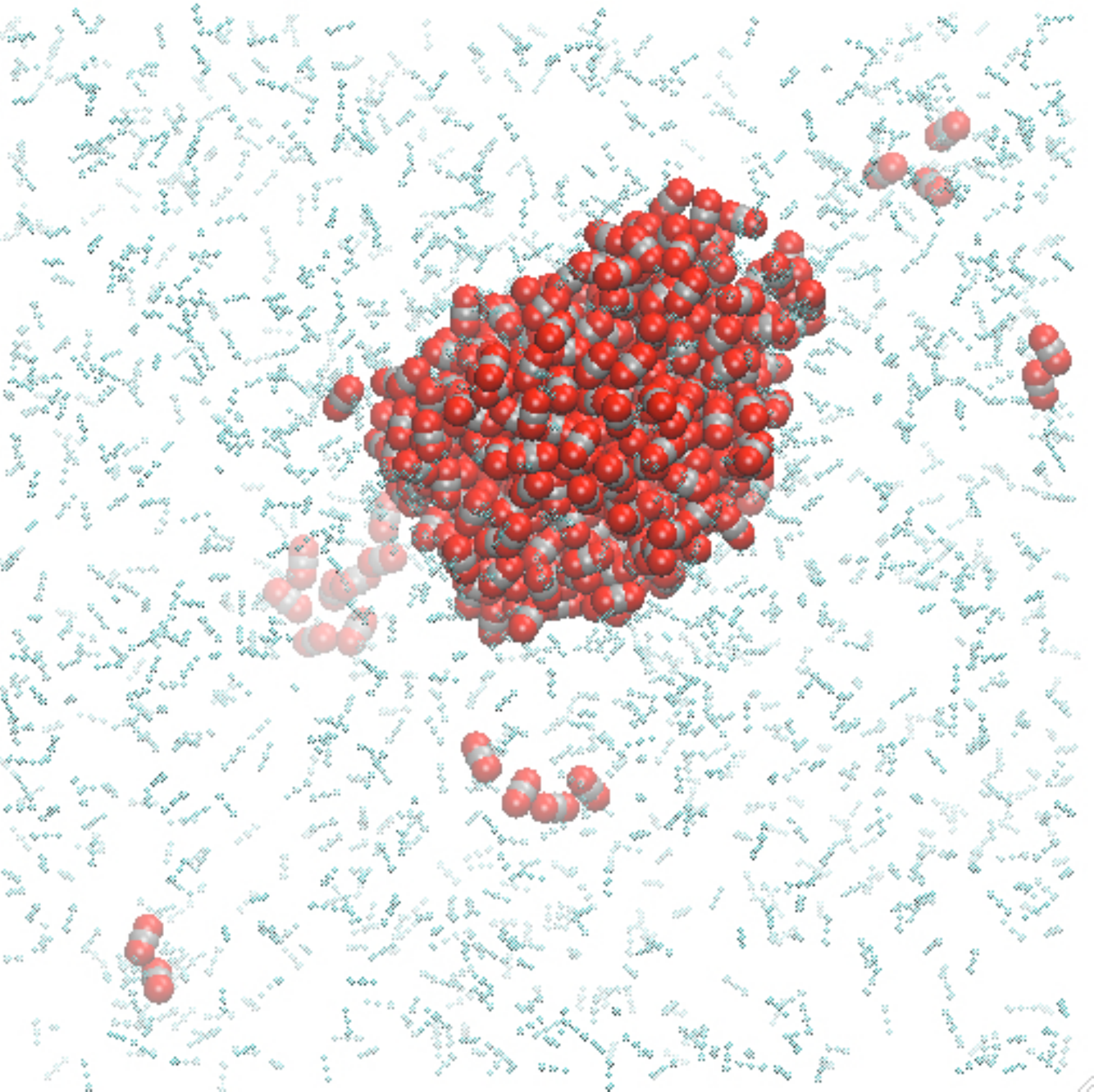}(c)
\end{center}
\caption{$CO_2$ at $T=260$~K: Snapshots of the system during the nucleation process for decreasing values of the entropy: (a) $S=3.66$~kJ/kg/K, (b) $S=3.6$~kJ/kg/K, (c) $S=3.54$~kJ/kg/K }
\label{Fig11}
\end{figure}

We now analyze the nucleation process in terms of the number of liquid-like molecules in the system. Closely following the process used to define this order parameter for the other systems, we determine the distributions for the number of neighbors in the supersaturated vapor and in the liquid using a cutoff distance of $5.85$~\AA~to characterize neighboring molecules. The distributions plotted in Fig.~\ref{Fig10}(a) allow us to distinguish between vapor-like molecules ($6$ neighbors or less) and liquid-like molecules ($7$ neighbors or more). Using this criterion, we determine the number of liquid-like molecules during the nucleation process as the entropy changes along the nucleation pathway. The results are plotted in Fig.~\ref{Fig10}(b) for system 7 and show a smooth increase in the number of liquid-like molecules as a function of the entropy during the formation of the liquid droplet. The formation of the liquid droplet can be best captured by looking at the snapshots of Fig.~\ref{Fig11}, which show the system for ddecreasing values of the entropy during the nucleation process. These snapshots show the increase in the size of the nucleus and reveal that the critical droplet contains $395\pm 33$ molecules for system 7. The critical sizes obtained for the other systems are the following: $290\pm 28$ molecules for system 8 and $242 \pm 20$ molecules for system 9. As for $Ar$ and $N_2$, the critical sizes are correlated with the free energy barriers of nucleation, with the highest barrier of system 7 corresponding to the biggest critical droplet and the lowest supersaturation.

\section{Conclusion}
In this work, we develop the $\mu VT-S$ simulation method to calculate the free energy of nucleation of atomic and molecular fluids along entropic pathways. We achieve this by working in the grand-canonical ensemble, which allows for the direct evaluation of entropy during the simulations and for the study of the nucleation process at a given supersaturation $\Delta \mu$, thereby providing a direct connection with the predictions from the classical nucleation theory. The simulation protocol  for $\mu VT-S$ consists of performing a series of umbrella sampling simulations that use $S$ as the reaction coordinate for the nucleation process. The results presented here show that the method provides a picture of the nucleation process that is consistent with the classical nucleation theory and with that found in previous work, including e.g. the dependence of the height of the free energy barrier and of the size of the critical nucleus on supersaturation. They also establish that the $\mu VT-S$ approach can be applied to atomic as well as molecular systems without any modification. Furthermore, our simulations reveal the correlation between the entropy of the system and the droplet size, and show how supersaturation impacts the entropic pathway visited during the nucleation process, with a narrowing of the range of entropy visited at high supersaturation. Our findings also allow us to identify the critical value for the entropy associated with the formation of a nucleus of a critical size. Further work presented in the next two papers of the series will focus on the development of the method for multi-component systems and for heterogeneous nucleation processes.

{\bf Acknowledgements}
Partial funding for this research was provided by NSF through CAREER award DMR-1052808.\\

\bibliography{muVTS_I}

\end{document}